%% file: ANA-EXOT-2025-06-PAPER.tex
\documentclass[PAPER, atlasdraft=false, UKenglish, cernpreprint, texmf, orcidlogo]{atlasdoc}
\usepackage{atlaspackage}
\usepackage{atlasbiblatex}

\usepackage{atlasphysics}
\graphicspath{{logos/}{figures/}}

\usepackage{ANA-EXOT-2025-06-PAPER-defs}

\AtlasTitle{Search for a leptoquark in events with a hadronically~decaying $\tau$-lepton and missing transverse momentum using $pp$ collisions at $\sqrt{s}=13~\TeV$ with the ATLAS detector}

\AtlasAbstract{%
A search for leptoquark signals is performed in final states with a hadronically decaying $\tau$-lepton and missing transverse momentum, using data from proton--proton collisions at a centre-of-mass energy of $\sqrt{s}=13$~TeV recorded by the ATLAS detector at the Large Hadron Collider during Run~2 (2015--18), corresponding to an integrated luminosity of 140~fb$^{-1}$.
The analysis is designed to probe both resonant production and non-resonant $t$-channel exchange of the leptoquark, covering a wide range of coupling scenarios.
No excess above the Standard Model background prediction is observed.
Limits are set on the couplings in the benchmark $U_1$ vector-leptoquark model at 95\% confidence level for masses between 1.5~TeV and 3.0~TeV.
}

\AtlasRefCode{EXOT-2025-06}

\PreprintIdNumber{CERN-EP-2026-136}

\arXivId{2606.02067}

\HEPDataRecord{3164547}

\AtlasJournal{\PLB 882 (2026) 141033}
\AtlasDOI{10.1016/j.physletb.2026.141033}

\hypersetup{pdftitle={ATLAS document},pdfauthor={The ATLAS Collaboration}}

\begin{document}

\maketitle

\section{Introduction}
\label{sec:intro}
%

Semileptonic decays of $B$ mesons provide one of the most sensitive probes of lepton flavour universality (LFU) in the Standard Model (SM) of particle physics, as well as possible contributions from physics \enquote{beyond the SM} (BSM).
In particular, a combination of the BaBar~\cite{BaBar:2012obs, BaBar:2013mob}, Belle~\cite{Belle:2015qfa, Belle:2016dyj, Belle:2017ilt, Belle:2019rba}, Belle-II~\cite{Belle-II:2024ami,Belle-II:2025yjp} and LHCb~\cite{LHCb:2015gmp, LHCb:2023zxo, LHCb:2023uiv,  LHCb:2024jll} results show a deviation of
about $3.8\sigma$ from the SM expectation using the HFLAV CKM2025 fit assumptions
for the ratio of branching fractions $\RD = \mathcal{B}(B \to D^{(*)}\tau\nu) / \mathcal{B}(B \to D^{(*)} \ell\nu)$,
where $\ell = e, \mu$ ($\ell = \mu$ for LHCb)~\cite{HFLAV:2019otj},
motivating scenarios with new interactions in the $b \to c\tau\nu$ transition.
Such extensions of the SM could be described by a leptoquark (LQ)~\cite{Pati:1974yy, Georgi:1974sy, Buchmuller:1986zs, Buchmuller:1986iq}, a hypothetical scalar or vector boson that couples directly to a lepton and a quark, thereby mediating interactions between the lepton and quark sectors.
The \RD\ anomaly can be accommodated by sizeable contributions from both the $b\tau$ and $c\nu$ couplings of a LQ~\cite{Buttazzo:2017ixm, Angelescu:2018tyl, Angelescu:2021lln}.

A comprehensive programme to search for LQ signals with $\tau$-leptons in the final state~\cite{ EXOT-2022-39, EXOT-2022-42, CMS-EXO-19-015, CMS-EXO-19-016, CMS-HIG-21-001, CMS-EXO-21-009correct} was developed by the ATLAS~\cite{PERF-2007-01} and CMS~\cite{CMS-TDR-08-001} experiments at the Large Hadron Collider (LHC).
In LQ pair production, the cross-section depends only on the strong coupling constant at leading order, making the searches largely insensitive to the specific choice of the LQ coupling.
Signals with LQ masses below approximately 1.6~\TeV\ were excluded by such searches~\cite{EXOT-2022-39, CMS-EXO-19-015}.
Searches for singly produced LQs that decay into $b\tau$ were performed in the $b\tau\tau$ final state, assuming they were produced in association with a $\tau$-lepton.
If the lepton--quark couplings are large, $t$-channel exchange of the LQ can lead to more events with a high-mass final-state than are expected in the SM.
ATLAS and CMS have performed such non-resonant searches for LQ signals in the $\tau\tau$ final state.
For the $U_1$ LQ model, the region with $\lambda_{b\tau}/\MLQ \gtrsim 1~\TeV^{-1}$ has been excluded by ATLAS~\cite{EXOT-2022-42}, where $\lambda_{b\tau} = g_U/\sqrt{2}$, with $g_U$ denoting the vector coupling constant and \MLQ\ the LQ mass, as defined in Section~\ref{sec:signal}.
CMS had similar sensitivity~\cite{CMS-EXO-19-016}, but observed a small excess of events at high $m_{\tau\tau}$, leading to a weaker observed limit on $\lambda_{b\tau}/\MLQ$.

\begin{figure}[bhtp]
\centering
\subfloat[]{
\includegraphics[width=0.4\linewidth]{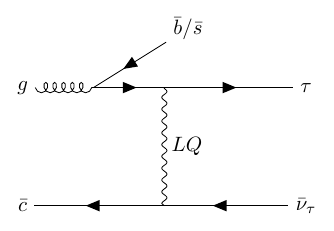}
}
\subfloat[]{
\includegraphics[width=0.5\linewidth]{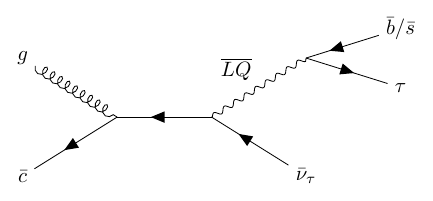}
}
\caption{Illustrative Feynman diagrams of signals targeted by this analysis: (a) non-resonant $t$-channel exchange of a leptoquark (LQ) and (b) a singly produced LQ decaying into $b\tau$/$s\tau$ (resonant signal).}
\label{fig:feynmanndiagram}
\end{figure}

The previous ATLAS and CMS searches in $\tau\tau(+b)$ final states mentioned above assumed that the signal rate depends only on the $b\tau$ coupling, because the $c\nu$ coupling contributes only through subleading effects. While a coupling involving a second-generation quark and a third-generation lepton is generally expected to be flavour-suppressed, a non-zero $c\nu$ coupling is required to accommodate the \RD\ anomaly. This provides a strong phenomenological motivation to probe this scenario directly with the first ATLAS analysis targeting the $\tau\nu(+b)$ final state, which is presented in this Letter.
Figure~\ref{fig:feynmanndiagram} shows illustrative Feynman diagrams for the signals targeted by this analysis.
A search for leptoquarks is performed in events with a hadronically decaying $\tau$-lepton (\tauh) with high transverse momentum ($\pt$) and large missing transverse momentum, whose magnitude is denoted by \met.
Two types of signal production modes are considered: a non-resonant process involving $t$-channel exchange of a LQ, typically associated with a low-\pt\ jet from the initial-state quark, and resonant single production of a LQ decaying into a high-\pt\ $b$- or $s$-quark and a $\tau$-lepton.
The analysis is further categorised according to the presence or absence of a jet containing $b$-hadrons (a \enquote{$b$-jet}).
The categories with a $b$-jet are sensitive to the product of the $b\tau$ and $c\nu$ couplings, directly relevant to the \RD\ anomaly, while categories without the $b$-jet requirement enhance sensitivity to scenarios with larger $c\nu$ coupling contributions.
An advantage of the $\tau\nu b$ channel is that the corresponding SM backgrounds are highly suppressed~\cite{Endo:2021lhi}.
In the SM, the production amplitudes of a $\tau\nu$ system with an associated $b$-quark,
which can directly interfere with the BSM signal, are CKM-suppressed.
Consequently, the SM-BSM interference is expected to be small in the high-mass region.
The CMS search of the $\tau\nu$ final state~\cite{CMS-EXO-21-009correct} included a LQ interpretation and was performed without $b$-jet requirements.

%
\section{ATLAS detector}
\label{sec:detector}
%

%
\newcommand{\AtlasCoordFootnote}{%
ATLAS uses a right-handed coordinate system with its origin at the nominal interaction point (IP)
in the centre of the detector and the \(z\)-axis along the beam pipe.
The \(x\)-axis points from the IP to the centre of the LHC ring, and the \(y\)-axis points upwards.
Polar coordinates \((r,\phi)\) are used in the transverse plane, \(\phi\) being the azimuthal angle around the \(z\)-axis.
The pseudorapidity is defined in terms of the polar angle \(\theta\) as \(\eta = -\ln \tan(\theta/2)\) and is equal to the rapidity
$ y = (1/2) \ln[(E + p_z)/(E - p_z)] $ in the relativistic limit.
Angular distance is measured in units of \(\Delta R \equiv \sqrt{(\Delta y)^{2} + (\Delta\phi)^{2}}\).}

The ATLAS detector~\cite{PERF-2007-01,GENR-2019-02} is a multipurpose apparatus with nearly \(4\pi\) coverage in
solid angle.\footnote{\AtlasCoordFootnote}
It consists of an inner tracking detector (ID) inside a \qty{2}{\tesla} superconducting solenoid, electromagnetic and hadronic calorimeters, and a muon spectrometer.
The inner tracking detector covers the pseudorapidity range \(|\eta| < 2.5\).
It consists of silicon pixel, silicon microstrip, and transition radiation tracking detectors.
Lead/liquid-argon (LAr) sampling calorimeters provide electromagnetic (EM) energy measurements
with high granularity within the region \(|\eta|< 3.2\).
A steel/scintillator-tile hadronic calorimeter covers the central pseudorapidity range (\(|\eta| < 1.7\)).
The endcap and forward regions are instrumented with LAr calorimeters
for EM and hadronic energy measurements up to \(|\eta| = 4.9\).
The muon spectrometer (MS) surrounds the calorimeters and is based on
three large superconducting air-core toroidal magnets with eight coils each.
The field integral of the toroids ranges between \num{2.0} and \qty{6.0}{\tesla\metre}
across most of the detector.
The MS includes a system of precision tracking chambers up to \(|\eta| = 2.7\) and fast detectors for triggering up to \(|\eta| = 2.4\).
A two-level trigger system is used to select events~\cite{TRIG-2016-01}.
The first-level trigger is implemented in hardware and uses a subset of the detector information
to accept events at a rate close to \qty{100}{\kHz}.
This is followed by a software-based trigger that
reduces the accepted rate of complete events to \qty{1.25}{\kHz} on average in Run~2,
depending on the data-taking conditions.
A software suite~\cite{SOFT-2022-02} is used in data simulation, in the reconstruction
and analysis of real and simulated data, in detector operations, and in the trigger and data acquisition systems of the experiment.


%
\section{Signal model}
\label{sec:signal}


This search focuses on the production of a vector leptoquark $U_1$ with quantum numbers $(\mathbf{3}, \mathbf{1}, {2/3})$, which correspond to the colour, weak, and hypercharge representations, respectively.
The effective Lagrangian for $U_{1}$ LQ interactions~\cite{Baker:2019sli} is
\begin{eqnarray*}
\mathcal{L}_{U_1} &=& -\frac{1}{2}U^\dagger_{1\mu\nu}U^{\mu\nu}_1+\MLQ^2 U^\dagger_{1\mu}U^{\mu}_1-ig_\text{s}(1-\kappa)U^\dagger_{1\mu}T^aU_{1\nu}G^{a\mu\nu} \nonumber \\
& & -ig_\text{Y}\frac{2}{3}(1-\tilde{\kappa}) U^\dagger_{1\mu}U_{1\nu}B^{\mu\nu} + \frac{g_{U}}{\sqrt{2}} \left[ U^\mu_1(\beta_\text{L}^{ij} \bar{q}_\text{L}^{i} \gamma_{\mu} \ell_\text{L}^{j} + \beta_\text{R}^{ij} \bar{d}_\text{R}^{i} \gamma_{\mu} e_\text{R}^{j})+h.c.\right], \label{eq:U1Model}
\end{eqnarray*}
where $U_{1\mu}$ is a vector LQ field, $U_{1\mu\nu} \equiv D_{\mu} U_{1\nu} - D_{\nu} U_{1\mu}$ is its field-strength tensor, \MLQ is the LQ mass, $G^{a}$ and $B$ denote the field-strength tensors of the SU(3)$_\text{c}$ and U(1)$_\text{Y}$ gauge bosons, $g_\text{s}$ and $g_\text{Y}$ the corresponding gauge coupling constants, and $T^a=\lambda^a/2$ (Gell-Mann matrices).
Here $q_\text{L}$ ($\ell_\text{L}$) denotes the left-handed quark (lepton) doublets and $d_\text{R}$ ($e_\text{R}$) denotes the right-handed down-type quark (charged-lepton) singlets,
$i \in \left\{1,2,3\right\}$ and
$j \in \left\{1,2,3\right\}$ are flavour indices,
and $a \in \left\{1,\ldots,8\right\}$ is a SU(3)$_\text{c}$ index.
The CKM matrix elements $V_{ji}$ are included in $q_\text{L}^{i}$ as
\begin{equation*}
q_\text{L}^i =
\begin{pmatrix}
V^*_{ji}u^j_\text{L} \\
d^i_\text{L} \\
\end{pmatrix}.
\end{equation*}
The flavour basis of the quark fields is identified with the mass basis of the down-type quarks.
The couplings $\beta_\text{L}$ and $\beta_\text{R}$ are complex $3 \times 3$ matrices in flavour space.
In this analysis, motivated by the discussion in Refs.~\cite{Cornella:2021sby, Aebischer:2022oqe}, they are defined as
\begin{equation*}
\beta_\text{L} = \left(
\begin{matrix}
0 & 0 & \beta_\text{L}^{13} \\
0 & 0 & \beta_\text{L}^{23} \\
0 & 0 & \beta_\text{L}^{33} \\
\end{matrix}
\right),
\hspace{0.5mm}
\beta_\text{R} = \left(
\begin{matrix}
0 & 0 & 0 \\
0 & 0 & 0 \\
0 & 0 & \beta_\text{R}^{33} \\
\end{matrix}
\right).
\end{equation*}
Conventionally, the normalisation factor $g_{U}$ is introduced to set $\beta_\text{L}^{33} = 1$.
There are two benchmark scenarios, corresponding to $\left|\beta_\text{R}^{33}\right|=0$ and $\left|\beta_\text{R}^{33}\right|=1$. %
In this analysis, the interpretation is performed in the purely left-handed scenario ($\left|\beta_\text{R}^{33}\right|=0$).
The parameters $\kappa$ and $\tilde{\kappa}$ introduce the LQ couplings to the SM gauge-boson fields.
They are set to $\kappa=\tilde{\kappa}=0$ in this analysis (Yang--Mills coupling scenario).

Integrating out the $U_1$ field at low energy induces semileptonic four-fermion operators.
The left-handed coefficient for the $b\to c\tau\nu$ transition scales as
\begin{equation}\label{eq:CLL}
C_\text{LL}^{c}
\sim
\frac{g_U^2 v^2}{4\MLQ^2}
\left(
1
+
\frac{V_{cs}}{V_{cb}}\beta_\text{L}^{23}
\right),
\end{equation}
where $v$ is the electroweak vacuum expectation value.
The term proportional to $\beta_\text{L}^{23}$ is enhanced by the  ratio $V_{cs}/V_{cb}$, implying that even moderate values of $\beta_\text{L}^{23}$ can generate sizeable contributions to the \RD\ anomalies.\footnote{The parameter
$\beta_\text{L}^{13}$ is set to $V_{td}^{*}/V_{ts}^{*}\ \beta_\text{L}^{23} \sim 0.2\ \beta_\text{L}^{23}$ in the signal generation, following the convention of Ref.~\cite{Baker:2019sli}.
Its contribution to Eq.~(\ref{eq:CLL}) is approximately 5\% of the $\beta_\text{L}^{23}$ corresponding term and is neglected there for simplicity.}
Although $\beta_\text{L}^{23} \sim 0.1$ is often assumed in UV-complete models~\cite{Cornella:2021sby, Aebischer:2022oqe}, low-energy charged-current observables alone do not impose stringent constraints on this parameter.
Additional constraints arise from flavour-changing neutral-current observables, particularly $B_{s(d)}$--$\bar{B}_{s(d)}$ mixing, as well as from LFU tests in $\tau$-lepton decays.
However, these bounds are model dependent and may be affected by additional states in the UV theory~\cite{Aebischer:2022oqe}.
This search therefore explores the high-$\beta_\text{L}^{23}$ region~\cite{Baker:2019sli}, which was not thoroughly covered by previous ATLAS and CMS searches.

The signal is primarily characterised by three parameters:
\MLQ, $g_U$ and $\beta_\text{L}^{23}$.
The relative contributions of resonant and non-resonant production depend strongly on \MLQ.
The resonant production cross-section decreases rapidly around $\MLQ \sim 1.5~\TeV$,
mainly due to the suppression of the parton distribution functions (PDFs) at high Bjorken-$x$.
As a result, the non-resonant contribution becomes dominant at higher masses.
The coupling $g_U$ controls the overall signal normalisation.
The parameter $\beta_\text{L}^{23}$ influences the flavour structure of the production process and changes the effective parton luminosity contributing to the signal.
Larger values of $\beta_\text{L}^{23}$ also reduce the relative fraction of signal events containing a $b$-quark in the final state.
The decay width of the LQ scales approximately as $\Gamma/\MLQ \propto g_U^2 \times \left[1 + (\beta_\text{L}^{23})^2\right]$.
For example, at $\MLQ = 1.5~\TeV$, $\beta_\text{L}^{23}=0.6$, and $g_U=2$, the intrinsic width is about 7\% of the LQ mass.
%
%
%


%
\section{Data and Monte Carlo simulation}
\label{sec:data_and_MC}


This search uses proton--proton ($pp$) collision data recorded at a centre-of-mass energy of $\sqrt{s}=13~\TeV$ by the ATLAS detector during Run~2 of the LHC (2015--2018),
corresponding to an integrated luminosity of 140~\ifb.
The number of interactions per bunch crossing (\pileup) ranged from about 8 to 70, with
an average of 34~\cite{DAPR-2018-01}.
Events were collected using the lowest-threshold unprescaled \met, single-electron and single-muon triggers~\cite{TRIG-2019-01, TRIG-2018-05, TRIG-2018-01}.
The typical thresholds were $\met = 110~\GeV$, with various \pileup mitigation algorithms depending on the data-taking period, and $\pt = 26~\GeV$ for electrons and muons, with different isolation requirements for various \pt\ ranges.
The $E_{\mathrm{T}}^{\mathrm{miss}}$ trigger is fully efficient above the offline $E_{\mathrm{T}}^{\mathrm{miss}}$ threshold used to select signal events. Therefore, no dedicated $\tau_{\mathrm{had}}$ trigger is required.

Signal LQ events with a $\tau$-lepton, neutrino, and up to two partons (including possible $b$-quarks)~\cite{Baker:2019sli} were modelled at leading order (LO) in QCD in the five-flavour scheme using the \MGNLO[3.3.1]~\cite{Alwall:2014hca} matrix element (ME) generator with the \NNPDF[3.0nlo] PDFs~\cite{Ball:2014uwa}.
The parton shower (PS) and hadronisation were generated by \PYTHIA[8.308]~\cite{Bierlich:2022pfr}, using parameter values from the A14 tune with  \NNPDF[2.3lo] PDFs~\cite{ATL-PHYS-PUB-2014-021} for the underlying event (UE).
All relevant production diagrams were considered in the ME calculation, including both the $s$-channel (resonant) and $t$-channel (non-resonant) topologies.
For the CKKW-L matching~\cite{Lonnblad:2011xx}, the Durham $k_{\perp}$ and $D$ parameters were set to 15~\GeV and 0.4, respectively, corresponding to a minimum jet \pt of 15~\GeV at the ME level.
Signal samples were generated for 152 different parameter points $(\MLQ, g_U, \beta_\text{L}^{23}$) satisfying $1.5 \leq \MLQ \leq 3.0~\TeV$, $0.5 \leq g_U \leq 3.0$ and $0 \leq \beta_\text{L}^{23} \leq 2.2$, for $\left|\beta_\text{R}^{33}\right|=0$.
The lower bound on \MLQ\ was chosen so as to cover the region around the exclusion limits set by previous LQ pair-production searches.
At higher masses ($\MLQ > 3~\TeV$), the signal is dominated by the non-resonant contribution, and both the cross-section and the kinematic distributions depend on the coupling-to-mass ratio rather than the explicit mass value. Therefore, 3.0~\TeV\ is taken as the upper boundary of the scan.
The contribution from interference between the signal process and the SM $\tau\nu$ processes with up to two partons was generated separately with \MADGRAPH at each parameter point, using the same configuration as for the signal process, and added to the corresponding signal sample.

The dominant source of SM background is $W\to\tau\nu$ production in association with a jet.
Production of a $W$ or $Z$ boson in association with jets ($W$+\;\!jets and $Z$\;\!+\;\!jets) was modelled using the \SHERPA[2.2.11] or 2.2.14 generator~\cite{Bothmann:2019yzt}, with
MEs calculated at next-to-leading-order (NLO) accuracy in QCD for up to two additional partons, and at LO
for up to five additional partons, and merged with the \SHERPA PS~\cite{Schumann:2007mg} using the \MEPSatNLO prescription~\cite{Hoeche:2011fd,Hoeche:2012yf,Catani:2001cc,Hoeche:2009rj}.
Top-quark pair (\ttbar) and single-top (single-$t$) backgrounds were generated at NLO accuracy in QCD with \POWHEGBOX[v2]~\cite{Frixione:2007nw,Nason:2004rx,Frixione:2007vw,Alioli:2010xd}, interfaced to \PYTHIA[8.230]~\cite{Sjostrand:2014zea} for parton showering and hadronisation. The \ttbar background was normalised to the inclusive cross-section calculated at next-to-next-to-leading order (NNLO) in QCD with resummation of next-to-next-to-leading logarithmic (NNLL) soft-gluon terms.
The single-$t$ contribution includes the $s$-channel, $t$-channel, and $Wt$-associated-production processes, with the $Wt$ channel providing the dominant contribution in this analysis.
Diboson processes were generated using \SHERPA[2.2.2] at NLO accuracy. Only the fully leptonic decay modes (i.e.\ $4\ell$, $\ell\ell\ell\nu$, $2\ell2\nu$, $\ell \nu\nu\nu$, and $4\nu$ where $\ell = e, \mu, \tau$) are considered in this analysis, as others contribute negligibly after the high-\met\ requirement.

The decays of bottom and charm hadrons were modelled with \EVTGEN~\cite{Lange:2001uf} for all samples except those generated with \SHERPA.
All generated event samples were passed through a detailed simulation of the ATLAS detector based on \GEANT~\cite{SOFT-2010-01, Agostinelli:2002hh}.
For the signal samples, the AtlFast3~\cite{SIMU-2018-04} tool based on generative neural networks provided a fast simulation of the calorimeter's response.
The effect of \pileup was modelled by overlaying~\cite{SIMU-2020-01} the simulated hard-scattering event
with inelastic proton--proton (\(pp\)) events generated from a mix of \EPOS[2.0.1.4]~\cite{Werner:2005jf} and \PYTHIA[8.308]~\cite{Bierlich:2022pfr}.
The \EPOS events were generated with the \EPOS LHC tune~\cite{Pierog:2013ria}
and the \PYTHIA events with the A3 tune~\cite{ATL-PHYS-PUB-2016-017}
and the \NNPDF[2.3lo]~\cite{Ball:2012cx} set of parton distribution functions (PDF).
\PYTHIA \pileup events include either a high transverse momentum (\pt) jet,
a prompt photon, or a lepton from a $b$-hadron decay,
while \EPOS was filtered to simulate all remaining \pileup events in the overlay sample.
The individual simulations were first reweighted to ensure a smooth connection across jet \pt then
the combination reweighted to match the distribution of the actual number of interactions per bunch crossing measured in data.

A summary of the MC generator configurations for all simulated event samples is shown in Table~\ref{tab:background_MC_list}.

\begin{table}[tbhp]
\caption{List of the MC simulated samples used in this analysis. The generators used for the ME calculation, the perturbative order of the ME calculation, the PS generator, the PDF set and the set of tuned parameters (tune) used for the UE modelling are shown. The symbol $\ell$ denotes an electron or muon.}%
\label{tab:background_MC_list}
\centering
\resizebox{\textwidth}{!}{
\begin{tabular}{llllll}
\toprule
Process & Generator & ME order & Parton shower & PDF set & Tune \\
\midrule
$U_1$ signal & \MGNLO & LO & \PYTHIA[8] & \NNPDF[3.0nlo] & A14 \\
\midrule
$W (\to \tau\nu)$+jets & \SHERPA[2.2.11] & NLO & \SHERPA & \NNPDF[3.0nnlo] & \SHERPA default\\
$\ttbar$ & \POWHEGBOX[v2] & NLO & \PYTHIA[8] & \NNPDF[3.0nlo] & A14 \\
Single-$t$ & \POWHEGBOX[v2] & NLO & \PYTHIA[8] & \NNPDF[3.0nlo] & A14 \\
$Z (\to \nu\nu)$+jets & \SHERPA[2.2.11] & NLO & \SHERPA & \NNPDF[3.0nnlo] & \SHERPA default\\
$Z (\to \tau\tau)$+jets & \SHERPA[2.2.14] & NLO & \SHERPA & \NNPDF[3.0nnlo] & \SHERPA default\\
$W (\to \ell\nu)$+jets & \SHERPA[2.2.11] & NLO & \SHERPA & \NNPDF[3.0nnlo] & \SHERPA default\\
$Z (\to \ell\ell)$+jets & \SHERPA[2.2.11] & NLO & \SHERPA & \NNPDF[3.0nnlo] & \SHERPA default\\
Diboson & \SHERPA[2.2.2] & NLO & \SHERPA & \NNPDF[3.0nnlo] & \SHERPA default\\
\bottomrule
\end{tabular}
}
\end{table}

%
%
%
%
%
%
%
%
%
%
%
%
%
%
%
%
%
%
%
%
%
%


%
\section{Event reconstruction and preselection}
\label{sec:object}


Hadronically decaying $\tau$-leptons are reconstructed from seed jets formed by the anti-$k_t$ algorithm~\cite{Cacciari:2008gp} with a radius parameter $R=0.4$, using topological cell-energy clusters, calibrated with the local hadronic cell-weighting scheme~\cite{PERF-2014-07, ATL-LARG-PUB-2009-001-2}.
All $\tauh$ candidates are required to have $\pt > 20~\GeV$ and $\left| \eta \right|<2.5$, excluding the EM calorimeter's barrel--endcap transition region $1.37 < \left|\eta\right| < 1.52$.
The vertex of a \tauh decay is determined from tracks within $\Delta R=0.2$ of the axis of the seed jet.
Geometric and impact-parameter criteria are then applied to all tracks associated with the seed jet to select those well matched to the \tauh candidate,
after which a track classifier based on a recurrent neural network (RNN)~\cite{ATL-PHYS-PUB-2022-044} distinguishes \tauh tracks from other tracks.
Each \tauh candidate is required to have one or three tracks classified as \tauh tracks, with the sum of their charges equal to $\pm 1$.
Another RNN-based classifier is used to separate $\tauh$ candidates from the hadron-jet background~\cite{ATL-PHYS-PUB-2022-044}, using tracks and calorimeter energy clusters associated with the seed jet, as well as several high-level discriminating variables.
Several different working points for \tauh identification, corresponding to different efficiencies, are used in this analysis.
The \enquote{Loose} working point is used for counting the number of $\tauh$, for the overlap removal described below and for \met\ reconstruction, whereas the \enquote{Tight} working point is employed for the \tauh in the signal selection.
In addition, all candidates are required to pass the \enquote{Loose} RNN eVeto working point~\cite{ATL-PHYS-PUB-2022-044}, to suppress electrons that are misidentified as \tauh candidates.

Electrons are reconstructed by combining energy clusters in the EM calorimeter with ID tracks~\cite{PERF-2017-01}.
Electron candidates are required to have $\pt>10~\GeV$ and $\left|\eta\right| < 2.47$ excluding the EM calorimeter's transition region.
All electrons are required to satisfy the \enquote{Tight} likelihood identification and \enquote{Loose\_VarRad} isolation criteria~\cite{EGAM-2018-01}.
Muons are reconstructed from ID and MS tracks, and additional calorimeter information~\cite{MUON-2018-03}.
Muon candidates are required to have $\pt > 10~\GeV$ and $\left|\eta\right|<2.5$.
All muons must satisfy the \enquote{Loose} identification working point and \enquote{Loose\_VarRad} isolation criterion~\cite{MUON-2018-03, MUON-2022-01}.
The \enquote{Loose} identification working point includes calorimeter-tagged muons, while the \enquote{Medium} working point includes only combined muons.
Both the electrons and muons are required to have small transverse ($d_0$) and longitudinal ($z_0$) impact parameters for their ID track, with $d_{0}/\sigma_{d_{0}} < 5$ (3) for electrons (muons) and $\left|z_{0}\sin\theta\right| < 0.5$~mm for both, where $\sigma_{d_{0}}$ is the uncertainty in the $d_0$ measurement.

Jets are reconstructed from particle-flow objects~\cite{PERF-2015-09} with the anti-$k_t$ algorithm ($R=0.4$), and are required to have $\pt > 20~\GeV$ and $\left| \eta \right|<2.5$.
\Pileup jets are suppressed using the jet--vertex tagger at the \enquote{FixedEffPt} working point~\cite{ATLAS-CONF-2014-018}.
Jets originating from $b$-hadrons (i.e.\ $b$-jets) are identified using the GN2 multivariate algorithm, based on a graph neural network and transformer, at a working point corresponding to 85\% $b$-jet efficiency~\cite{FTAG-2023-05}.

To avoid double-counting, overlapping physics objects are rejected in the following order:
a \tauh within $\Delta R = 0.2$ of an electron; $\tauh$ candidates within $\Delta R = 0.2$ of a muon;
calorimeter-tagged muons sharing an ID track with an electron;
electrons sharing an ID track with a muon;
the closest jet within $\Delta R = 0.2$ of an electron;
electrons within $\Delta R= 0.4$ of a jet;
jets within $\Delta R = 0.4$ of a muon if they have at most two associated tracks and the muon is ghost-associated with the jet;
muons within $\Delta R = 0.4$ of a jet;
and jets within $\Delta R = 0.2$ of a \tauh.
All the reconstructed physics objects above are calibrated and the corresponding efficiency scale factors and energy/momentum corrections are applied to simulated events to account for residual differences between data and simulation.

The missing transverse momentum is defined as the negative vector sum of the \pt\ of all selected objects in the event, plus a term to account for the \pt\ of soft particles with ID tracks that are matched to the primary vertex but not to any of the selected objects~\cite{JETM-2020-03}.

Only data from periods with stable beams and all detector systems operational are employed~\cite{DAPR-2018-01}.
The reconstructed interaction point with the highest $\sum \pt^2$ of the associated tracks with $\pt > 500~\MeV$ is selected as the primary vertex~\cite{ATL-PHYS-PUB-2015-026}.
All events must have at least two tracks associated with the primary vertex.
Events must contain at least one jet.
To suppress multijet background, events are rejected if the smallest azimuthal separation between jets and \met\ (i.e.\ $\min \Delta \phi(j, \met))$ is less than 0.4 and the ratio of \met\ to jet \pt\ for the jet with the smallest $\Delta \phi(j, \met)$ is less than 6.
The requirements above define the event preselection, which serves as the starting point for the subsequent event categorisation and statistical analysis.


%

%
\section{Event selection}
\label{sec:selection}


Four signal regions (SRs) are defined in order to maximise the sensitivity of the analysis: two for non-resonant LQ production and two for resonant LQ production,  illustrated by the Feynman diagrams shown in the left and right panels of Figure~\ref{fig:feynmanndiagram}, respectively.
Each region is further subdivided according to the presence or absence of a $b$-jet, as summarised in Table~\ref{tab:SR_EventSelection}.
During the optimisation of the SRs, using simulated events, the selection thresholds were chosen by maximising the expected exclusion sensitivity, while requiring
the MC statistical uncertainty in the expected background to remain below 30\%.
The jet multiplicity requirements suppress top-quark background, while the remaining kinematic requirements target the resonant or non-resonant signal topology.
Events in the SRs are required to pass the \met\ trigger.
Exactly one \tauh with $\pt > 200~\GeV$ is required.
Events with additional light leptons (electrons or muons) are vetoed.

Resonance and non-resonance SRs are primarily distinguished by the jet \pt.
The resonance SRs, \VRW-\Res and \SR-\Res, with no or one $b$-jet in each event, are optimised for signals with lower LQ masses ($\sim$1.5~\TeV).
Events must satisfy $\met > 200~\GeV$ and $\mt > 200$~\GeV, where $\mt$ is the transverse mass of the \tauh and $\met$.
Up to four jets are allowed per event.
In \VRW-\Res, the number of $b$-jets must be zero, and the leading jet's \pt\ must exceed 250~\GeV.
The leptoquark mass is reconstructed as the invariant mass of the leading jet and the \tauh, denoted by $m_{j\tauh}$, which is required to be greater than 800~\GeV.
In \SR-\Res, exactly one $b$-jet is required, and its $\pt$ must exceed 250~\GeV.
In addition, the invariant mass of the $b$-jet and the \tauh, denoted by $m_{b\tauh}$, is required to be greater than 800~\GeV.

The non-resonance SRs, \VRW-\NonRes and \SR-\NonRes, target the high-\mt\ regime with either no or one $b$-jet, and hence $\mt>600~\GeV$ and $\met > 400~\GeV$ are required.
The azimuthal separation between the \tauh and \met, denoted by \DphiTauMet, is required to be greater than 1.2.
To suppress top-quark background, at most two jets are allowed per event.
In \VRW-\NonRes, the leading jet's \pt\ is required to satisfy $50 < \pt < 250~\GeV$, while for \SR-\NonRes, the $b$-jet's \pt is used instead.

\begin{table}[ht]
\caption{Summary of event selection criteria in SRs. All events are required to satisfy the preselection requirements.}
\label{tab:SR_EventSelection}
\centering
\begin{tabular}{lcccc}
\toprule
Feature & \multicolumn{4}{c}{Criterion} \\
\midrule
Trigger requirement & \multicolumn{4}{c}{\met trigger (lowest-unprescaled)} \\
\tauh multiplicity ($N_{\tauh}$) & \multicolumn{4}{c}{1} \\
$\pt^{\tauh}$ [$\GeV$] & \multicolumn{4}{c}{$>200$} \\
\tauh identification & \multicolumn{4}{c}{Tight} \\
Light lepton multiplicity ($N_{\ell}$) & \multicolumn{4}{c}{0}  \\
\hline\hline
& \multicolumn{4}{c}{\hspace{-1cm}Region} \\
& \VRW-\Res & \SR-\Res & \VRW-\NonRes & \SR-\NonRes \\
\midrule
$b$-jet multiplicity ($N_{b}$) & 0 & 1 & 0 & 1  \\
$b$-jet $\pt$ [$\GeV$] & -- & $>250$ & -- & $[50, 250]$\\
Leading jet $\pt$ [$\GeV$] & $>250$ & -- & [50, 250] & -- \\
\met[$\GeV$] & \multicolumn{2}{c}{$>200$} & \multicolumn{2}{c}{$>400$} \\
\mt[$\GeV$] & \multicolumn{2}{c}{$>200$} & \multicolumn{2}{c}{$>600$} \\
$m_{b\tauh}$ [$\GeV$] & -- & $>800$ & -- & -- \\
$m_{j\tauh}$ [$\GeV$] & $>800$ & -- & -- & -- \\
\DphiTauMet & \multicolumn{2}{c}{--} & \multicolumn{2}{c}{$>1.2$} \\
Number of jets & \multicolumn{2}{c}{$\leq4$} & \multicolumn{2}{c}{$\leq2$} \\
\bottomrule
\end{tabular}
\end{table}


%

%
\section{Background estimation}
\label{sec:backgound}

The SM backgrounds are estimated using simulated samples.
The dominant SM background in the SRs comes from $W \to \tau\nu$ production in association with jets.
Subdominant contributions arise from top-quark decays producing $\tau$-leptons ($\ttbar \to \tau\nu b\tau\nu b$ and $Wt \to \tau\nu \tau\nu b$), in cases where $\tau$-leptons and $b$-quarks are misreconstructed as light-flavour jets.
The normalisations of $W$+\;\!jets and top-quark samples are constrained by the data in high-purity control regions (CRs), where single-lepton triggers are used instead of the \met\ trigger.
Events with a jet misidentified as a \tauh, mostly $Z\to\nu\nu$ events with large \met, make minor contributions in this analysis, and are also estimated from simulation with a conservative uncertainty obtained from data-driven studies with $Z\to\ell\ell$ events.
Multijet background is found to be negligible after the preselection.
Other SM backgrounds, including diboson and $Z \to \tau\tau$ contributions, as well as minor contributions from $W \to \ell\nu$ and $Z \to \ell\ell$ processes, are also estimated by simulation.

The $W$+\;\!jets background normalisation is constrained using two control regions,
CRW-Res and CRW-NonRes, constructed using the same event selections as in the resonant and non-resonant SR0b regions, respectively, except that an electron or
muon is required instead of a \tauh. They achieve a purity of approximately
70\% in $W\to\ell\nu$ events.
The electrons and muons selected in the control regions are required to satisfy tighter identification and/or isolation criteria than those used in the region definition and rejection of overlapping events, which are defined in Section~\ref{sec:object}.
Electrons (muons) in the $\CRW$ regions must satisfy the \enquote{TightLH} (\enquote{Medium})
identification and \enquote{HighPtCaloOnly} (\enquote{Tight\_VarRad}) isolation criteria~\cite{EGAM-2018-01,MUON-2022-01}.
Events with $b$-jets are vetoed to suppress top-quark contributions.
All other selection criteria are kept
the same as in the corresponding SR0b regions, except that \tauh-related
kinematic variables are computed using the electron or muon instead of the \tauh.

The top-quark background normalisation is constrained using two control regions,
CRTop-Res and CRTop-NonRes, defined by requiring one \tauh and one additional electron or muon.
To increase the event yield in these regions, several kinematic requirements are loosened with respect to the SR selections:
$N_{b} \geq 1$, $\pt^{\tauh}>100~\GeV$, $\met>100~\GeV$, $\mt>100~\GeV$, and no restriction on $m_{b\tauh}$.
Among multiple $b$-jets, the one with the highest \pt is selected and used as a proxy for the $b$-jet in the \SR regions, and is employed in the reconstruction of kinematic variables such as $m_{b\tauh}$.
All other selection criteria are kept the same as in the \SR\ regions.
A purity of approximately 90\% in top-quark events is achieved.

The background normalisation estimated in the CRs is tested in dedicated validation regions (VRs), designed to validate the extrapolation from the CRs to the SRs.
These include regions with modified $b$-jet multiplicity or light-lepton requirements, while having kinematic selections similar to those in the SRs.
Two types of VRs are defined, for each of the resonance and non-resonance cases, amounting to four VRs in total. The VRs are labelled as
\VRb and \VRTop, characterised by the requirements $(N_{\tauh}, N_{\ell}, N_{b}) = (0, 1, 1)$ and $(1, 0, 2)$, respectively.
The \VRb regions are used to test the extrapolation of the background estimation from the CRs with a $b$-jet veto to the SRs requiring a $b$-jet.
In the \VRb regions, similarly to the \CRW ones, the kinematic variables used in the event selection are computed with an electron or a muon, instead of a \tauh.
The \VRTop regions probe the validity of the \ttbar\ background normalisations obtained in CRs using a $\tauh$ and an electron or muon, by testing them in regions with a $\tauh$ and no light lepton.
In the \VRTop regions, the highest-\pt $b$-jet is used as a proxy of the $b$-jet in the SRs.
As with the \CRTop regions, looser selection cuts for $N_{b}$, $\pt^{\tauh}$, $\met$, $\mt$ and $m_{b\tauh}$ are required.


%

%
\section{Systematic uncertainties}
\label{sec:systematics}


Experimental and modelling systematic uncertainties are included as normalisation and shape variations in the fit and are smaller than the statistical uncertainty from the limited SR event yields.

The uncertainty in the combined 2015--2018 integrated luminosity is 0.83\%~\cite{DAPR-2021-01}, obtained using the LUCID-2 detector~\cite{LUCID2} for the primary luminosity measurements, complemented by measurements using the inner detector and calorimeters.

The $\tau$-lepton energy scale and its uncertainties are measured in situ using $Z \to \tau\tau$ events, combined with test beam and MC simulation comparison in the high-\pt\ region~\cite{ATL-PHYS-PUB-2015-045, ATLAS-CONF-2017-029, ATL-PHYS-PUB-2022-044}.
The correction applied to the simulated $\tau$ energy scale is typically below 1\% of the reconstructed $\tauh$ energy.
The associated uncertainty in the absolute $\tau$ energy scale is smaller than 1.5\%.
The efficiencies of $\tau$-lepton reconstruction and identification in the simulation samples are corrected to agree with data, depending on \pt, using \ttbar events.
The central value of the correction factor is close to unity in all \pt\ ranges, with uncertainties of 5\%--8\% (6\%--11\%) for 1-prong (3-prong) \tauh\ at the \enquote{Tight} identification working point.
The uncertainties in the electron and muon reconstruction and identification efficiencies and their energy and momentum scales and resolutions are considered,  and they are significantly smaller than those affecting $\tau$-leptons.

The jet energy scale and resolution and \met\ soft-term uncertainties are estimated using in situ methods~\cite{JETM-2018-05, JETM-2020-03}.
The efficiency of $b$-jet identification in the simulation is corrected to match that measured in \ttbar\ data events, and uncertainties in the correction factor are estimated~\cite{FTAG-2018-01,FTAG-2020-08, FTAG-2019-02}. %
For signal samples using AtlFast3 detector simulation, an additional uncertainty taking into account how it differs from the full \GEANT simulation is considered, and its impact on the signal yield is around 1\%.

Uncertainties in the theoretical modelling of simulated events are also considered.
For simulated $W$+\;\!jets events,
variations of the renormalisation and factorisation scales, by factors of 2 and $1/2$, and the PDFs are considered~\cite{Ball:2022oua}.
Additionally, approximate NLO electroweak corrections are applied to the nominal $W$+\;\!jets samples,
and the corresponding effect is taken as an additional systematic uncertainty.
Since the normalisation of $W \to \tau\nu$ events is estimated in the \CRW regions with a $b$-jet veto and applied to SRs with a $b$-jet, a 30\% uncertainty is assigned in those regions to the ratio of $W$-boson production in association with heavy-flavour hadrons ($W$+\;\!HF) to that with light-flavour hadrons ($W$+\;\!LF), according to previous ATLAS studies~\cite{ATL-PHYS-PUB-2017-006, PMGR-2021-01, STDM-2018-43, HIGG-2020-20}.
The $W$+\;\!HF modelling uncertainty is decorrelated between the resonant and non-resonant regions.

For simulated top-quark events,
variations of the ME calculation, renormalisation and factorisation scales, and PDFs are considered, similarly to the $W$+\;\!jets samples, along with variations related to initial- and final-state radiation and the $h_{\textrm{damp}}$ parameter~\cite{ATL-PHYS-PUB-2020-023}. %
The difference in the top-quark \pt\ distribution's shape between the nominal NLO prediction and the NNLO prediction~\cite{Czakon:2017wor} is also considered as an additional uncertainty.
Uncertainties in the interference between \ttbar\ and single-$t$ events are estimated by comparing simulations using the diagram removal and diagram subtraction schemes~\cite{Frixione:2008yi}. %

The background containing a jet misidentified as a $\tauh$ is a minor contribution in the SRs and is mainly due to $Z \to \nu\nu$ production.
The modelling of the probability of misidentifying a jet as a $\tauh$ in simulated events is checked in a control sample requiring $Z \to \ell\ell$ decays, together with a $\tauh$ candidate, providing a sample with more than 90\% purity in events containing $Z$ boson and a jet misidentified as $\tauh$ candidate.
The kinematic distributions are generally well modelled by the simulation, while up to 60\% differences are observed in the overall event yield between data and simulation.
These differences include effects from both the modelling of $Z$+\;\!jets and the probability for jets to be misidentified as $\tauh$ candidates, with the latter expected to dominate.
Since this background contributes only a small fraction of the total background in the SRs,
a conservative 100\% uncertainty is assigned to the normalisation of those simulated events.
For the signal samples, variations of the renormalisation and factorisation scales and PDF are taken into account.


%

%
\section{Results}
\label{sec:result}


The \textsc{TRExFitter} framework~\cite{aly_2025_17790864} is used to perform the statistical analysis.
Two fit configurations are employed: one is the \enquote{background-only} fit and the other is the \enquote{signal-plus-background (s+b)} fit.
A template maximum-likelihood fit~\cite{Cowan:2010js} is performed using the observed event yields in each region.
For the background-only fit, the four single-bin CRs are included in the fit to constrain the normalisations of $W$+\;\!jets and top-quark samples, with a fixed signal strength of $\mu = 0$.
Separate background normalisation factors are introduced for the resonance and non-resonance regions, resulting in four independent normalisation factors: two for $W$+\;\!jets ($\lambda_{W}^{\Res}$ and $\lambda_{W}^{\NonRes}$) and two for the top-quark background ($\lambda_{t}^{\Res}$ and $\lambda_{t}^{\NonRes}$). The top-quark normalisation factors are applied simultaneously to both the \ttbar\ and single-$t$ samples.
The fit results are applied to SRs and VRs to compare the observed data with the SM expectation.
For the s+b fit, on the other hand, a fit is performed in the four CR and four SR event categories simultaneously to constrain the background normalisation and determine the signal strength $\mu$.
The expected yield is parameterised in the likelihood as $\mu S+\sqrt{\mu}I+B$, where $S$ denotes the pure BSM contribution, $I$ denotes the interference between the BSM and SM $\tau\nu$ amplitudes, and $B$ denotes the SM background contributions.
The interference term scales with $\sqrt{\mu}$, since the pure BSM contribution is quadratic while the interference is linear in the BSM amplitude.
In the SRs with a $b$-jet veto (\VRW-\Res and \VRW-\NonRes), the expected contribution from destructive interference for $\mu=1$ amounts to approximately 20\% of the purely BSM yield at $\MLQ=1.5~\TeV$, increasing to about 30\%--40\% at higher masses.
The impact of interference on the signal yield is negligibly small in the SRs with a $b$-jet (\SR-\Res and \SR-\NonRes).
It is verified that, for the range of coupling values considered in this analysis, the total BSM signal plus interference yield remains positive in all cases.
Nuisance parameters are correlated across analysis regions and between signal and background samples, where applicable, unless stated otherwise.
The sensitivity is driven by the SR1b regions. However, for larger values of $\beta_\text{L}^{23}$ a sizable part of the signal populates regions without a reconstructed $b$-jet and the SR0b regions are sensitive to such $U_1$ signals.

For both fit configurations, systematic uncertainties are considered as nuisance parameters in the fit, with Gaussian, log-normal or Poisson constraints.
Statistical uncertainties in each bin due to the limited size of the simulated background samples are taken into account by dedicated parameters using the Beeston--Barlow \enquote{lite} (one parameter per bin) technique~\cite{Barlow:1993dm}.

The background-only fit result is shown in Figure~\ref{fig:unblinded_Bonly_fit_summary}.
Figure~\ref{fig:CRVR_dataMC_comparison} shows the  kinematic distributions in \CRW\ and \VRb regions after the background-only fit.
The best-fit normalisation factors are $\lambda_{W}^{\Res} = 1.21 \pm 0.11$, $\lambda_{W}^{\NonRes} = 0.89 \pm 0.08$, $\lambda_{t}^{\Res}=0.76 \pm 0.16$ and $\lambda_{t}^{\NonRes}=0.81\pm 0.16$.
The post-fit background predictions agree with data within approximately $1\sigma$ in all VRs except in the VRW-NonRes region, where a deviation of about $2\sigma$ from the SM expectation is observed.
No indications of specific detector malfunctions are found after a verification of the events in this region.
The background modelling in this region is further validated by checking the level of agreement between
data and prediction in regions of lower and higher \met using the same event selection, and no evidence
of significant mismodelling or deficiencies is found.
Given the size of the deviation, the observation is consistent with a statistical fluctuation.

In the \VRW-\Res and \VRW-\NonRes regions, 33 and 45 events are observed, with expected backgrounds of $36.4\pm7.3$ and $38.3\pm9.9$ events, respectively.
In the \SR-\Res and \SR-\NonRes regions, 3 and 2 events are observed, compared to background expectations of $2.69 \pm 0.83$ and $1.88 \pm 0.71$ events, respectively.
Figure~\ref{fig:unblinded_SR_N_minus_1} shows the kinematic distributions in each SR after the background-only fit.
Overall, the observed data are in good agreement with the SM prediction.

\begin{figure} [htpb]
\centering
\includegraphics[width=0.7\linewidth]{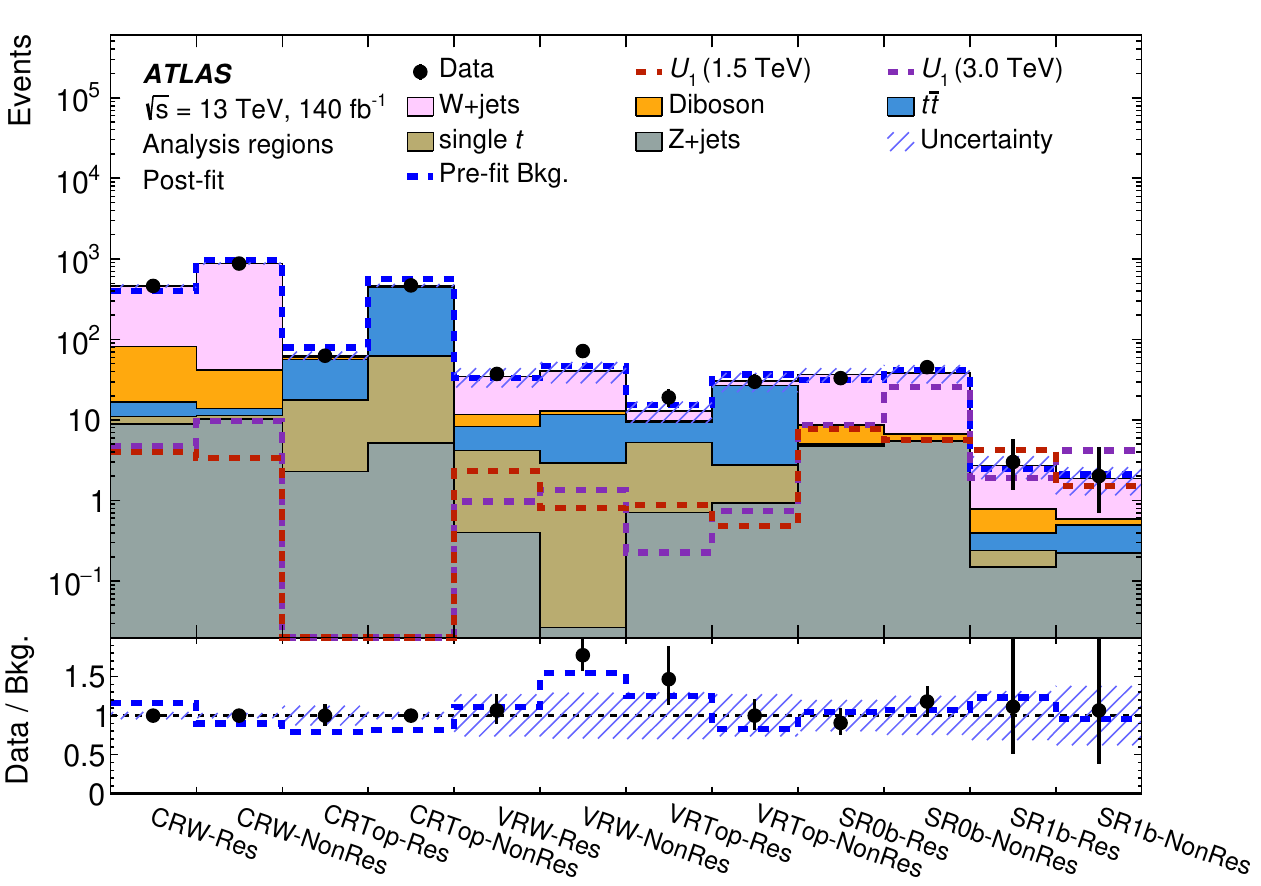}
\caption{Comparison of the data and predicted background event yields for four CRs, four VRs and four SRs in the upper panel. The background prediction is shown after a \enquote{background-only fit} in the CRs (extrapolated to the VRs and SRs). The signal expectation for signal strength $\mu=1$ is shown for illustrative purposes at the parameter points $\left(\MLQ, g_{U}, \beta_\text{L}^{23}\right) = (1.5~\TeV, 1.5, 0.6)$ and $(3.0~\TeV, 2.5, 1.0)$ with a red dashed line and a violet dashed line, respectively. The ratio of the data to the background (\enquote{Bkg.}) prediction is shown in the lower panel, separately for post-fit background (black points) and pre-fit background (dashed blue line). The size of the combined statistical and systematic uncertainty in the background prediction is indicated by a blue hatched band.}
\label{fig:unblinded_Bonly_fit_summary}
\end{figure}

\begin{figure} [htpb]
\centering
\subfloat[]{\includegraphics[width=0.45\linewidth]{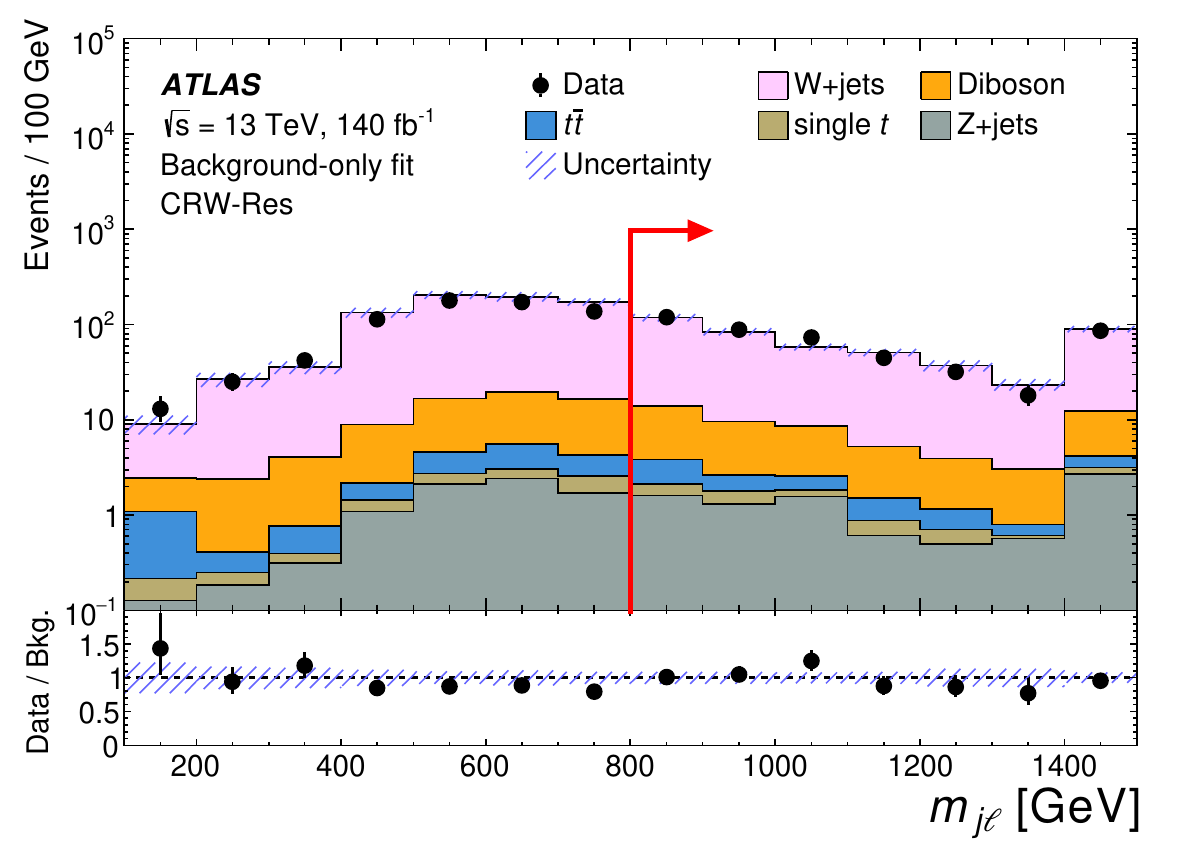}}
\subfloat[]{\includegraphics[width=0.45\linewidth]{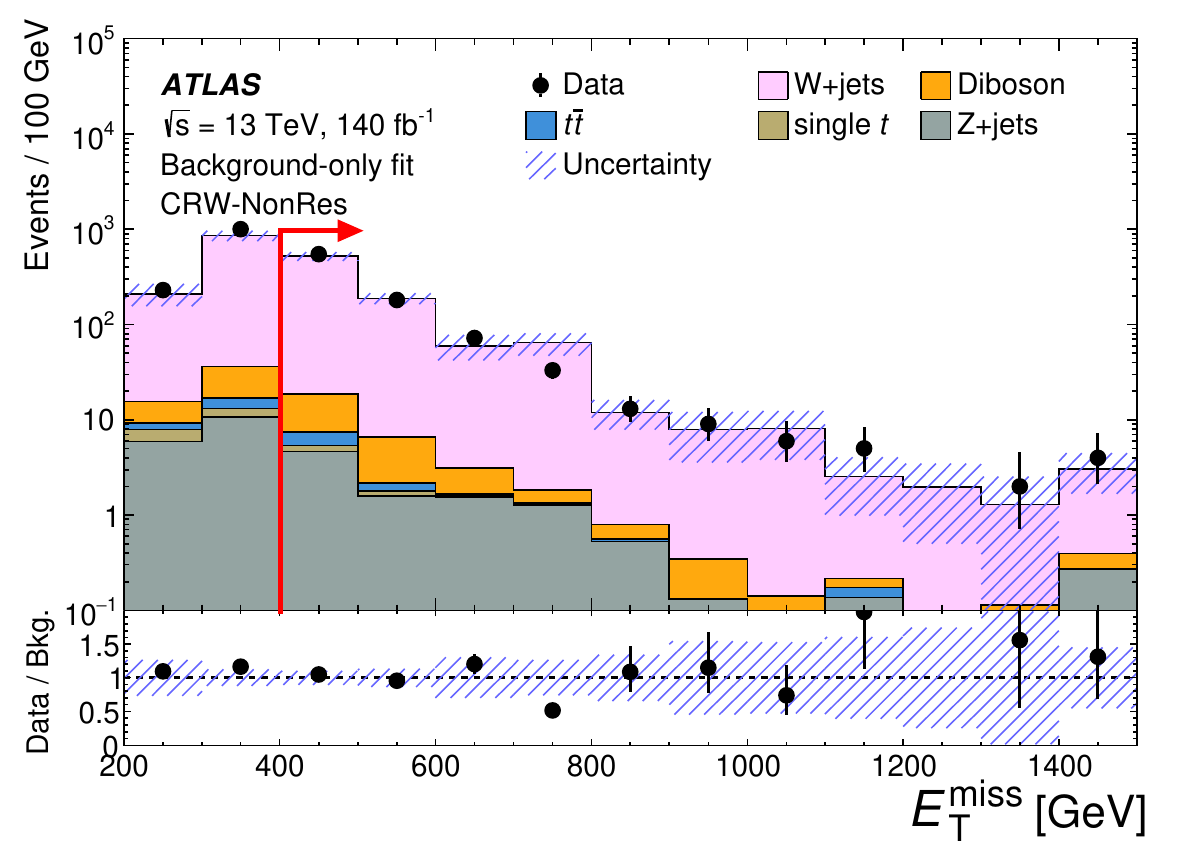}}\\
\subfloat[]{\includegraphics[width=0.45\linewidth]{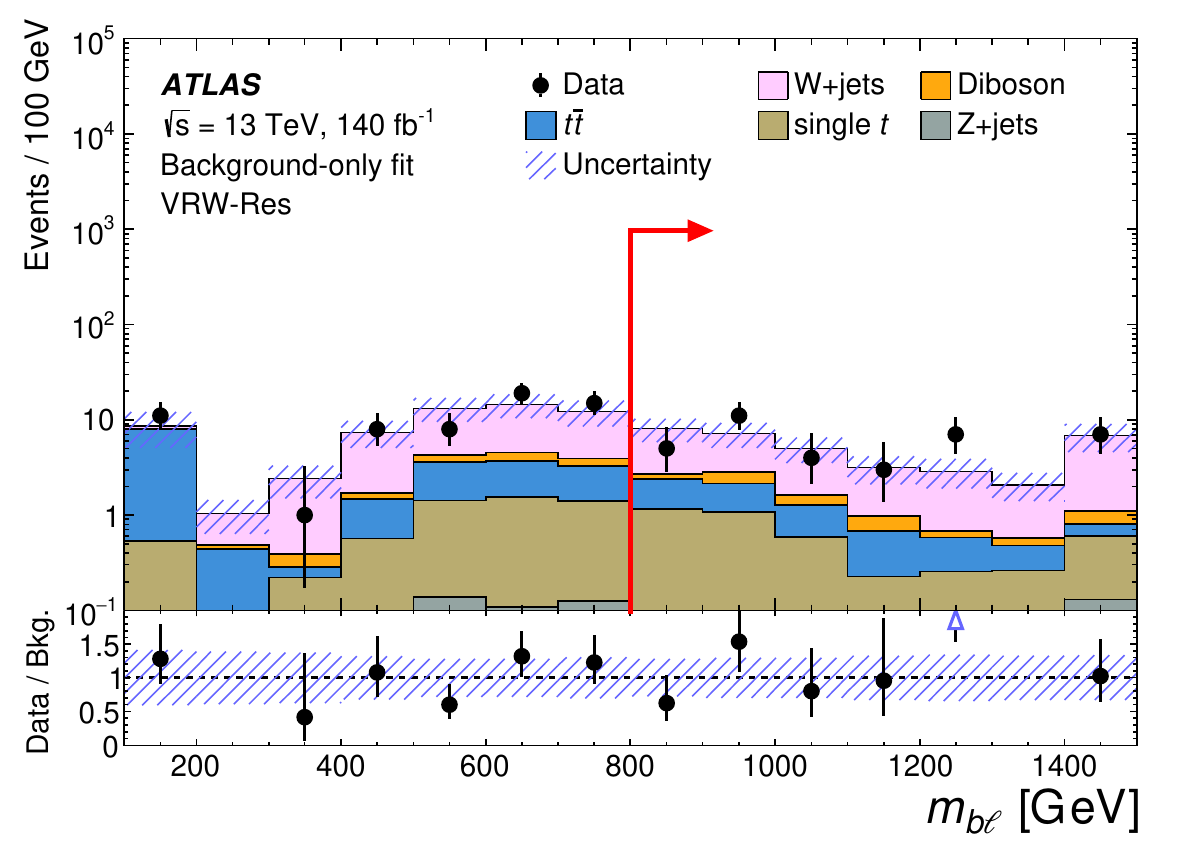}}
\subfloat[]{\includegraphics[width=0.45\linewidth]{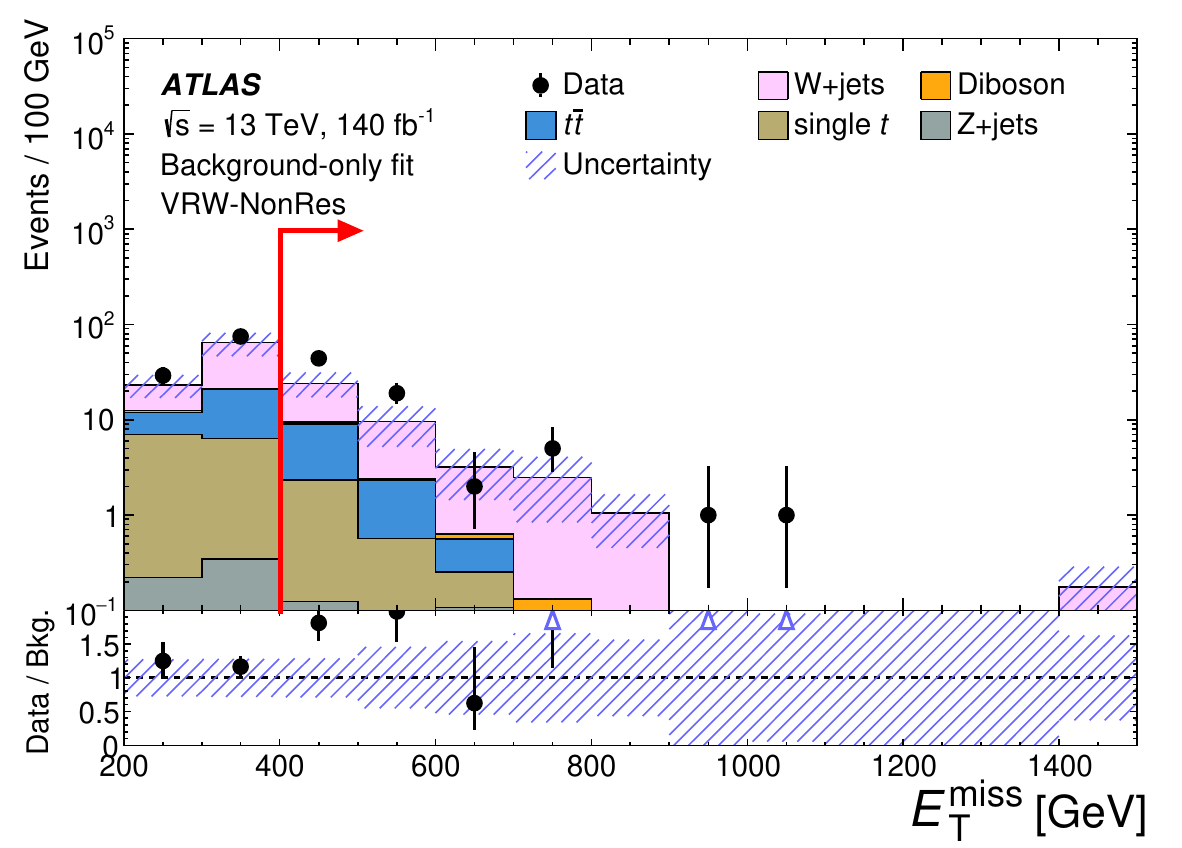}}
\caption{The post-fit distributions of (a) $m_{j\ell}$ in \CRW-\Res, (b) $\met$ in \CRW-\NonRes,
(c) $m_{b\ell}$ in \VRb-\Res and (d) $\met$ in \VRb-\NonRes.
The overflow entries are added to the rightmost bin.
The background-only fit result is used for the background prediction.
The distributions are shown prior to imposing the selection requirement on each variable, which is indicated by a right-pointing red arrow.
Bottom panels show the ratio of the observed data to the background (`Bkg.') prediction. Blue upward triangles indicate ratio values exceeding the displayed y-axis range.
The size of the combined statistical and systematic uncertainty in the background prediction is indicated by a blue hatched band.}
\label{fig:CRVR_dataMC_comparison}
\end{figure}

\begin{figure}[p]
\centering
\subfloat[]{\includegraphics[width=0.45\linewidth]{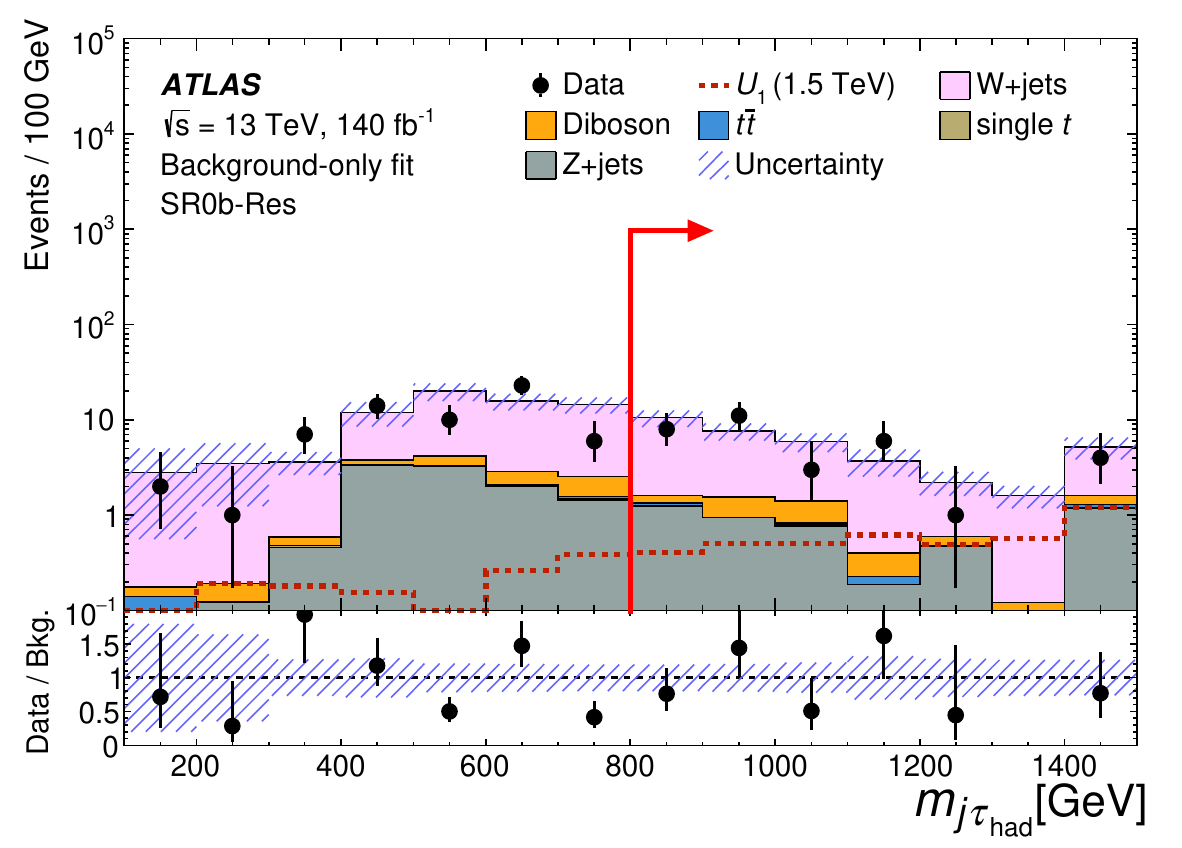}}
\subfloat[]{\includegraphics[width=0.45\linewidth]{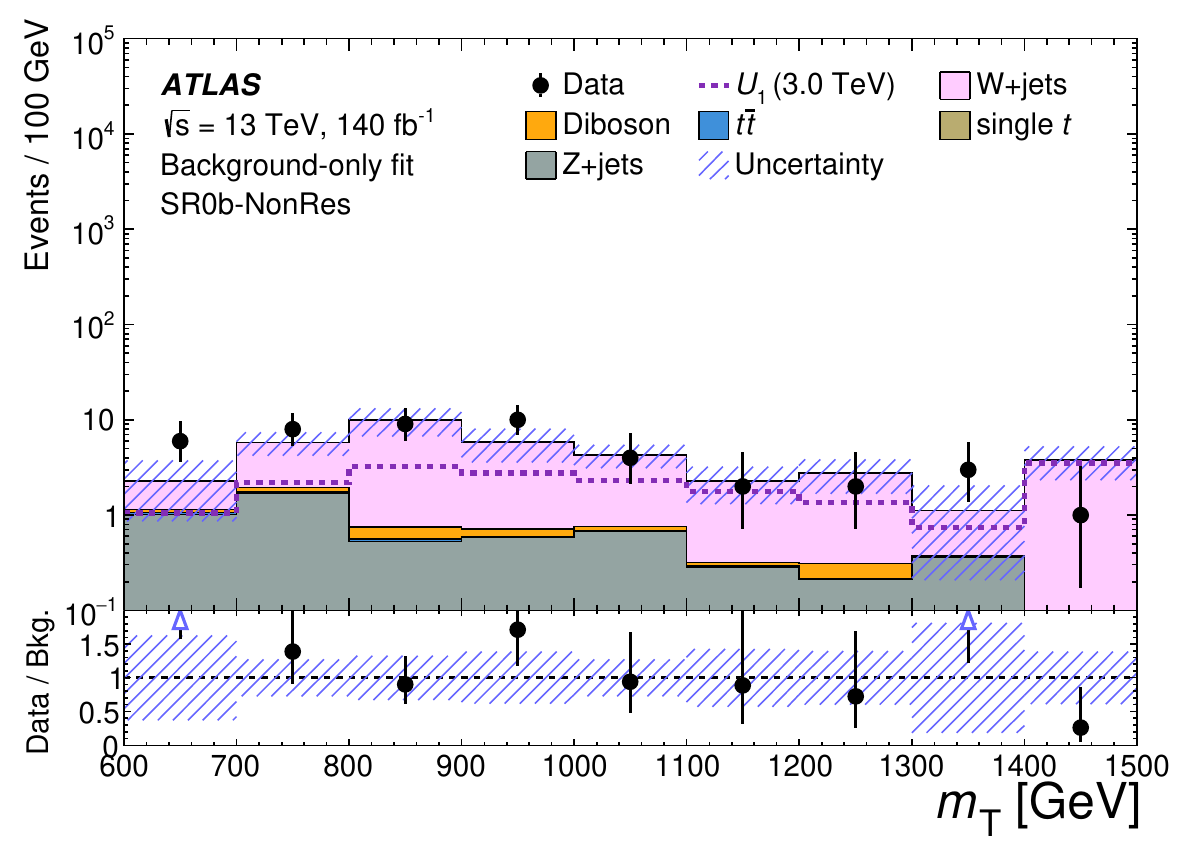}}\\
\subfloat[]{\includegraphics[width=0.45\linewidth]{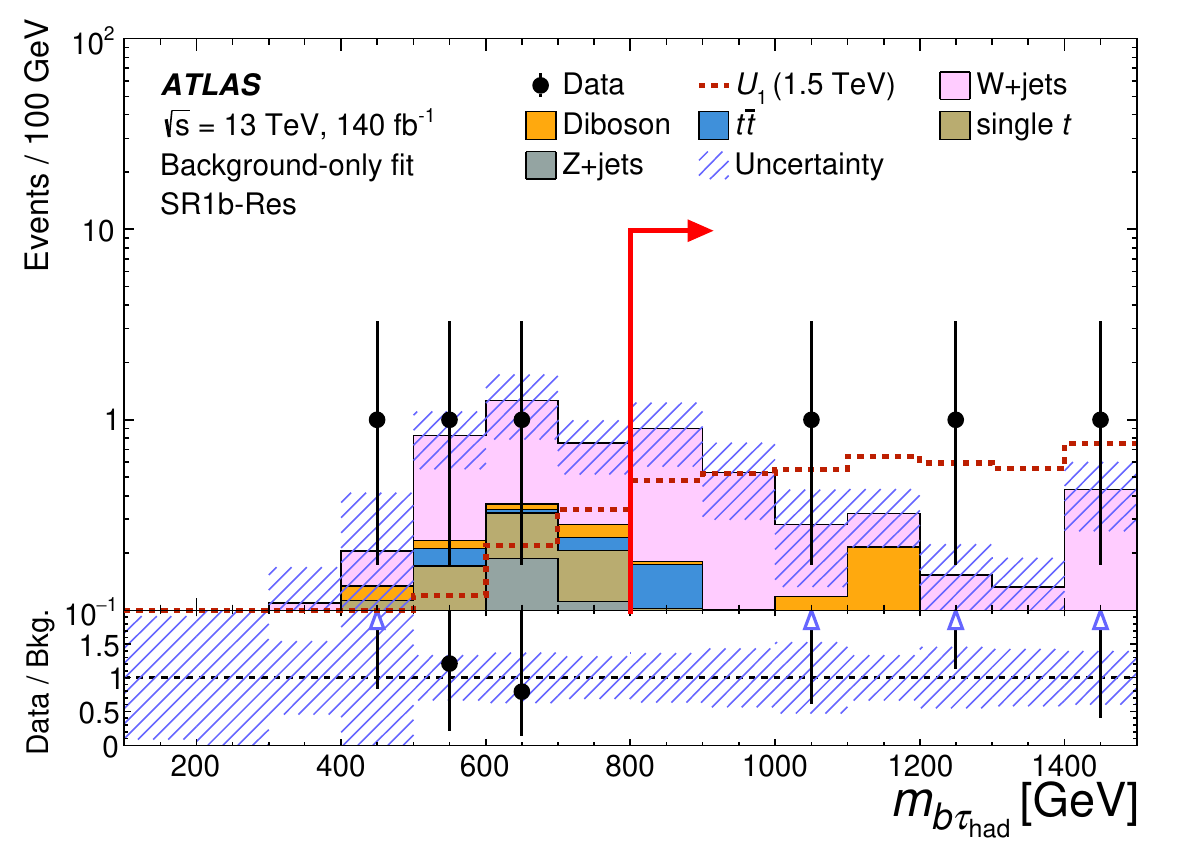}}
\subfloat[]{\includegraphics[width=0.45\linewidth]{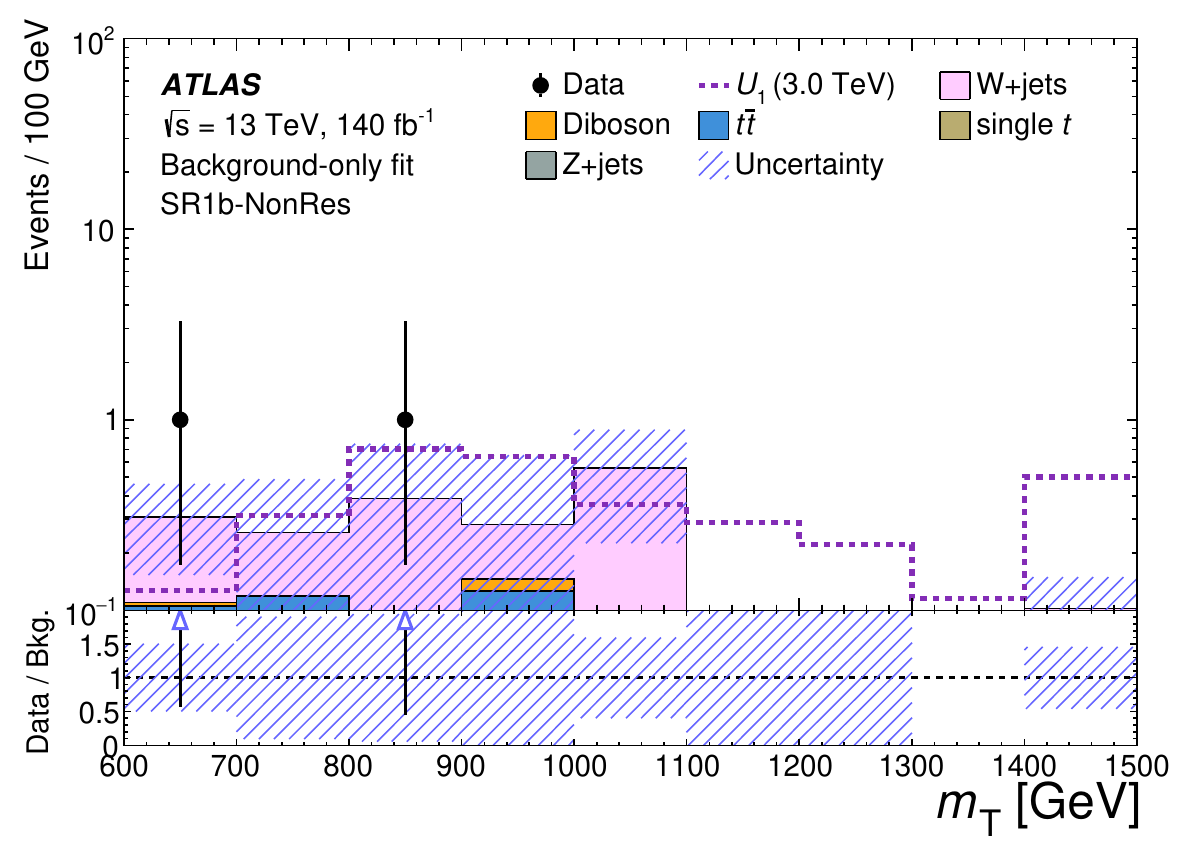}}
\caption{The post-fit distributions of (a) $m_{j\tauh}$ in \bVetoSR-\Res,
(b) $\mt$ in \bVetoSR-\NonRes,
(c) $m_{b\tauh}$ in \SR-\Res, and (d) $\mt$ in \SR-\NonRes.
The overflow entries are added to the rightmost bin.
The background-only fit result is used for the background prediction. The signal expectation at the parameter points $\left(\MLQ, g_{U}, \beta_\text{L}^{23}\right) = (1.5~\TeV, 1.5, 0.6)$ and $(3.0~\TeV, 2.5, 1.0)$ is also shown with a red dashed line and a violet dashed line, respectively. The distributions are shown prior to imposing the selection requirement on each variable, which is indicated by a right-pointing red arrow.
Bottom panels show the ratio of the observed data to the background (`Bkg.') prediction. Blue upward triangles indicate ratio values exceeding the displayed y-axis range.
The size of the combined statistical and systematic uncertainty in the background prediction is indicated by a blue hatched band.}
\label{fig:unblinded_SR_N_minus_1}
\end{figure}

The impact of each systematic uncertainty on the analysis sensitivity is evaluated with respect to the best-fit value of the signal strength $\mu$.
This evaluation is performed using a dataset from pseudo-datasets constructed from the post-fit background expectation with an injected signal corresponding to $\mu=1$, and the impact is assessed on the fitted value of $\mu$.
Using the results of the s+b fit, performed with the dataset from pseudo-experiments in the SRs and the observed data in the CRs, the impact of a given nuisance parameter is determined from the covariance matrix, assuming a linear propagation of uncertainties around the best-fit point.
For each nuisance parameter, its contribution to the uncertainty in the signal strength is estimated as the product of its post-fit uncertainty, the total post-fit uncertainty in $\mu$, and the correlation coefficient between the nuisance parameter and the signal strength.
The dominant source of systematic uncertainty is the $\tau$-lepton energy scale, and its impact on $\mu$ is around 10\%.
The uncertainty in the ratio of $W$+\;\!HF to $W$+\;\!LF is the subdominant source of systematic uncertainty, whose impact depends on \MLQ, but typically amounts to about 5\%--10\%.

An upper limit on the signal strength $\mu$ is calculated with the CL$_\text{s}$ method~\cite{Read:2002hq}, where the distributions of the test statistic are evaluated using pseudo-experiments, at predefined points in the $(\MLQ, g_U, \beta_\text{L}^{23})$ parameter space.
Figures~\ref{fig:unblinded_limit_gu_mu} and \ref{fig:unblined_limit_beta_gu} show the observed and expected exclusion contours at 95\% confidence level (CL) in the purely left-handed scenario ($|\beta_\text{L}^{33}|=1, |\beta_\text{R}^{33}|=0$).
Figure~\ref{fig:unblinded_limit_gu_mu} presents the limits on $g_{U}$ as a function of $\MLQ$ at fixed values of $\beta_\text{L}^{23}$, while Figure~\ref{fig:unblined_limit_beta_gu} shows the limits on $\beta_\text{L}^{23}$ as a function of $g_{U}$ for each $\MLQ$ point.
The observed (expected) upper limits on $g_{U}$ are 1.13 (1.11) at $\beta_\text{L}^{23}=1.0$ and 0.52 (0.53) at $\beta_\text{L}^{23}=2.2$ for a leptoquark mass of 1.5~\TeV, and 2.30 (2.19) at $\beta_\text{L}^{23}=1.0$ and 1.22 (1.18) at $\beta_\text{L}^{23}=2.2$ for a leptoquark mass of 2.5~\TeV.
The shaded areas indicate the parameter region consistent with the \RD\ anomalies~\cite{Aebischer:2022oqe}.
In Ref.~\cite{Aebischer:2022oqe}, $\beta_\text{L}^{23}$ is constrained to be within the interval $[0.06, 0.16]$ with the general UV-completion assumption, but in this Letter, it is mapped to a higher $\beta_\text{L}^{23}$ region using Eq.~(\ref{eq:CLL}). %
These limits are complementary to those from LQ pair-production searches~\cite{EXOT-2022-42,CMS-EXO-19-016}, which are largely insensitive to $g_{U}$ at leading order and therefore provide the strongest constraints at low masses, whereas the present $\tau\nu(+b)$ analysis probes the high-$\beta_\text{L}^{23}$ region through single-production and non-resonant contributions.
A previous $\tau\tau$ analysis~\cite{EXOT-2022-42}, which assumed $\beta_\text{L}^{23}\sim\mathcal{O}(0.1)$, excluded values of approximately $g_U>2.1$, $2.8$, and $3.5$ at $\MLQ=1.5~\TeV$, $2.0~\TeV$ and $2.5~\TeV$, respectively.
In contrast, this analysis extends the sensitivity to a complementary region of the parameter space, characterised by lower values of $g_{U}$ and higher values of $\beta_\text{L}^{23}$.

\begin{figure}[htbp]
\centering
\subfloat[]{\includegraphics[width=0.45\linewidth]{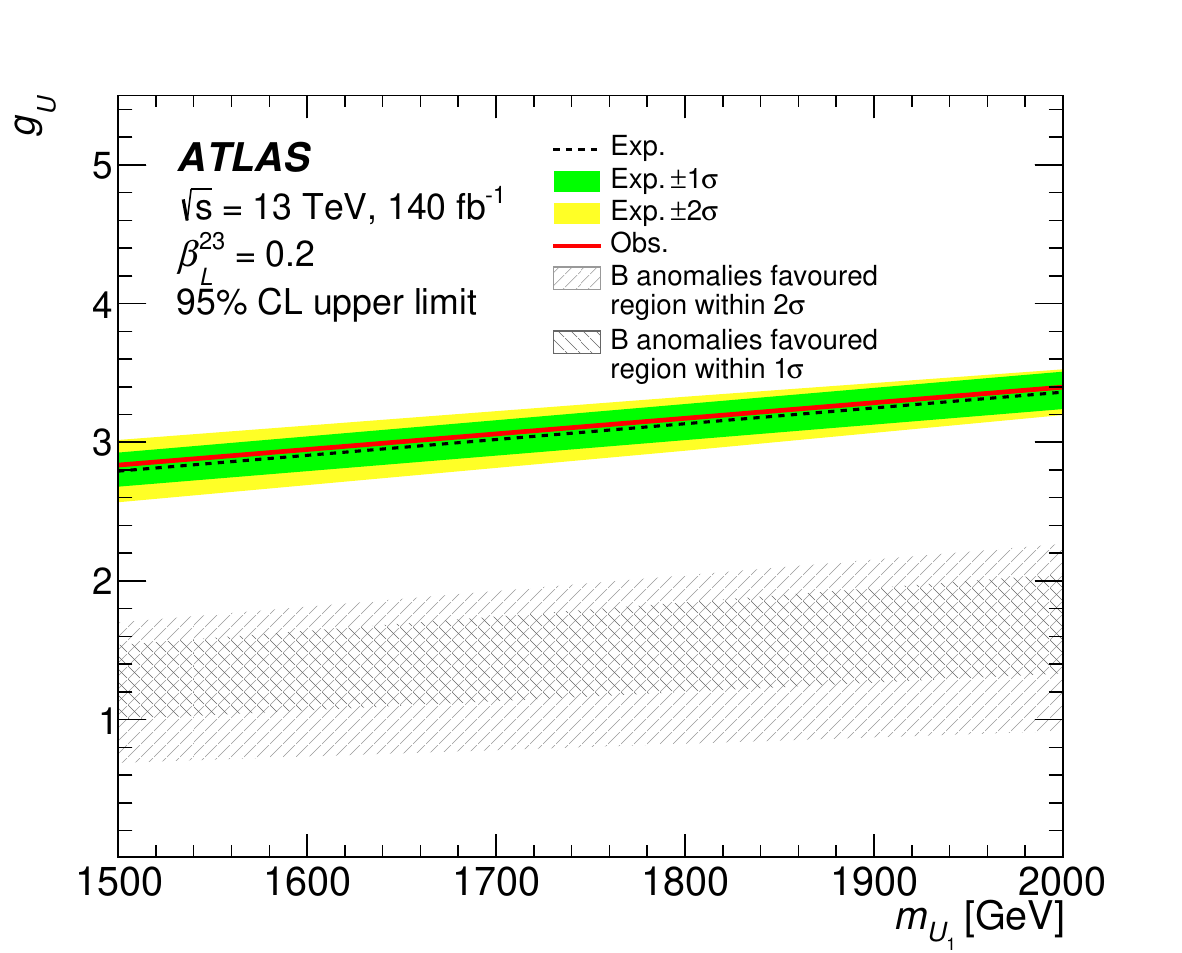}} \\
\subfloat[]{\includegraphics[width=0.45\linewidth]{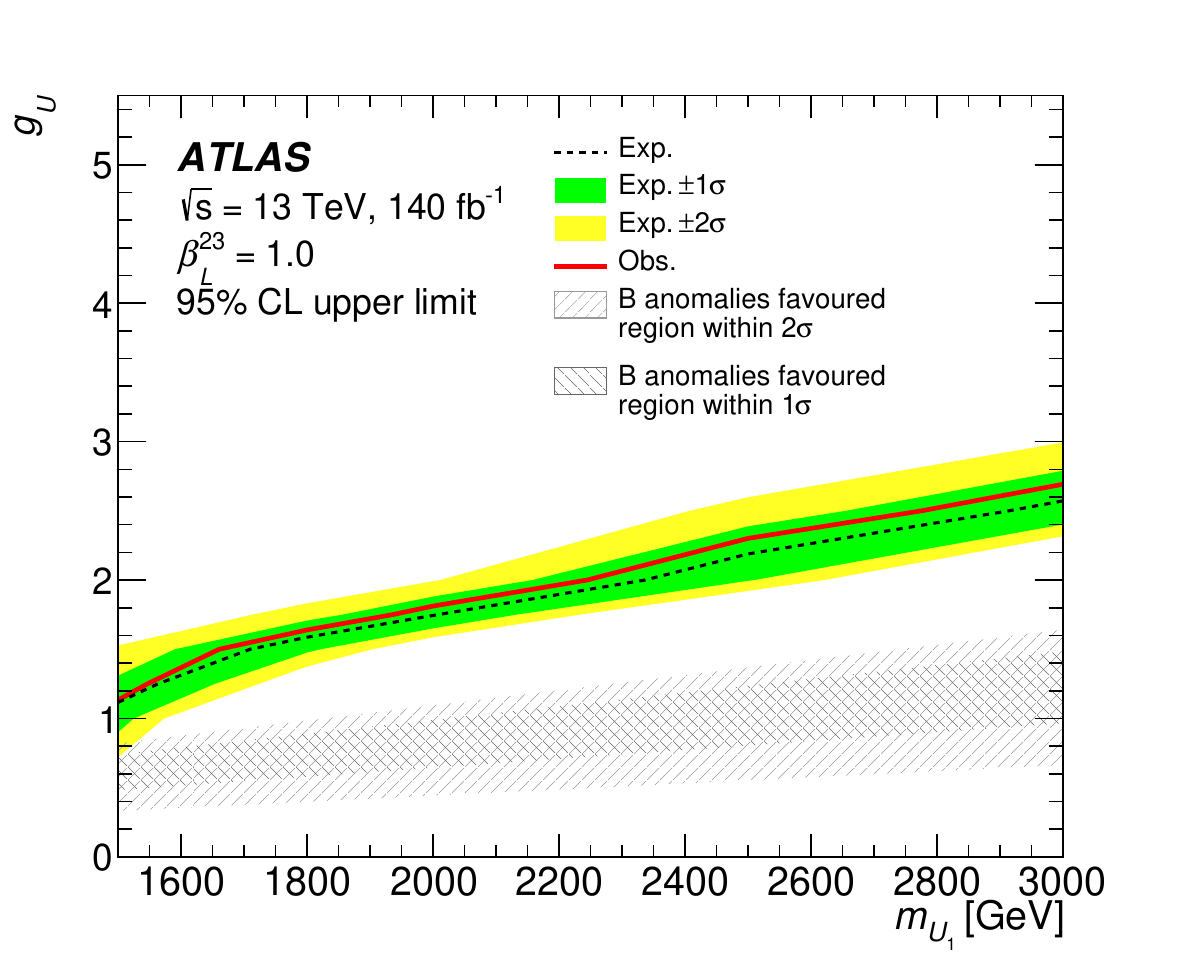}}
\subfloat[]{\includegraphics[width=0.45\linewidth]{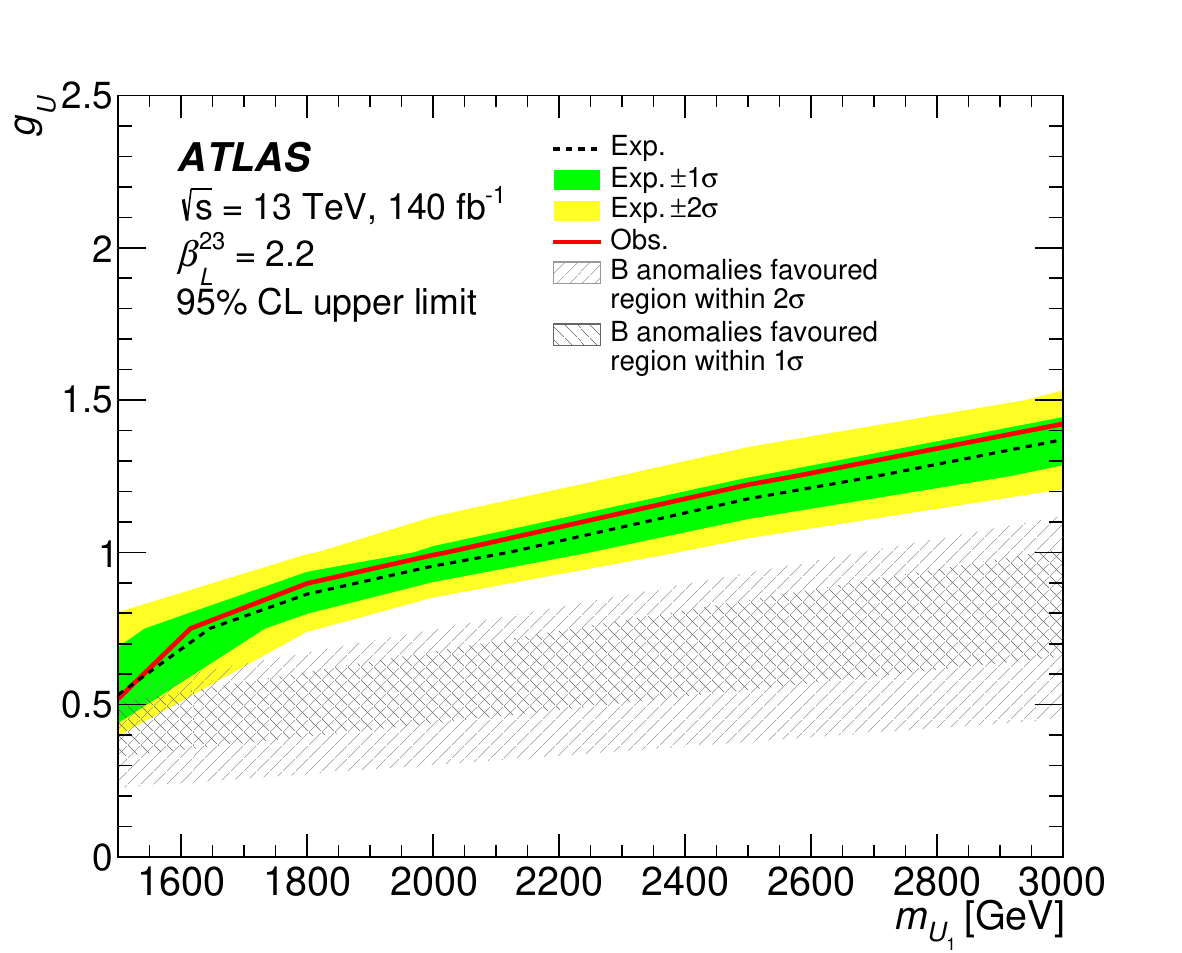}} \\
\caption{Observed (red full line) and expected (black dashed line) 95\% CL upper limits on $g_U$ as a function of $\MLQ$, at (a) $\beta_\text{L}^{23}=0.2$, (b) 1.0 and (c) 2.2. The shaded regions are favoured by $B$ anomalies at the 1$\sigma$ and 2$\sigma$ levels, converted from $C_\text{LL}^{c}$ limit in Ref.~\cite{Aebischer:2022oqe} using Eq.~(\ref{eq:CLL}).}
\label{fig:unblinded_limit_gu_mu}
\end{figure}

\begin{figure}[htbp]
\centering
\subfloat[]{\includegraphics[width=0.45\linewidth]{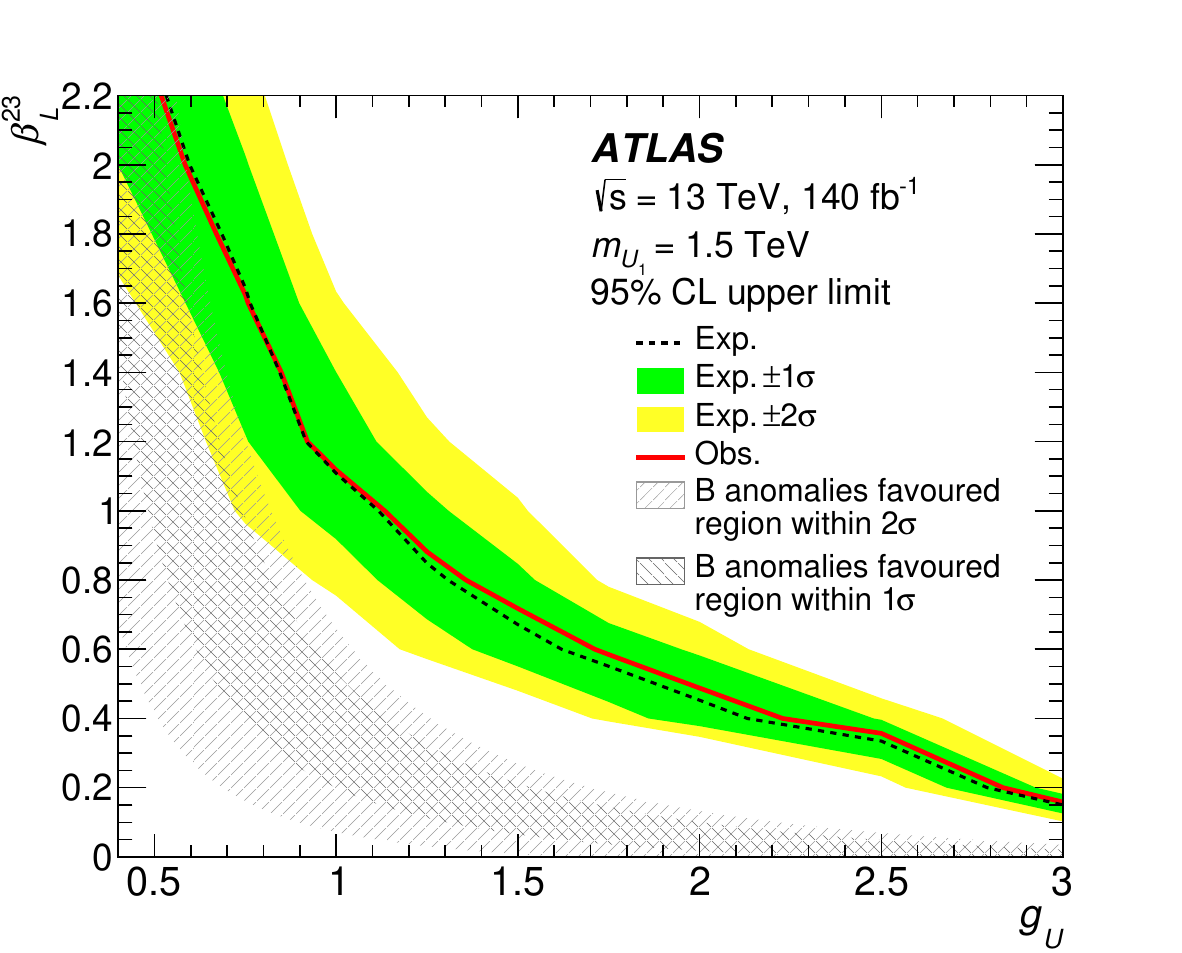}}
\subfloat[]{\includegraphics[width=0.45\linewidth]{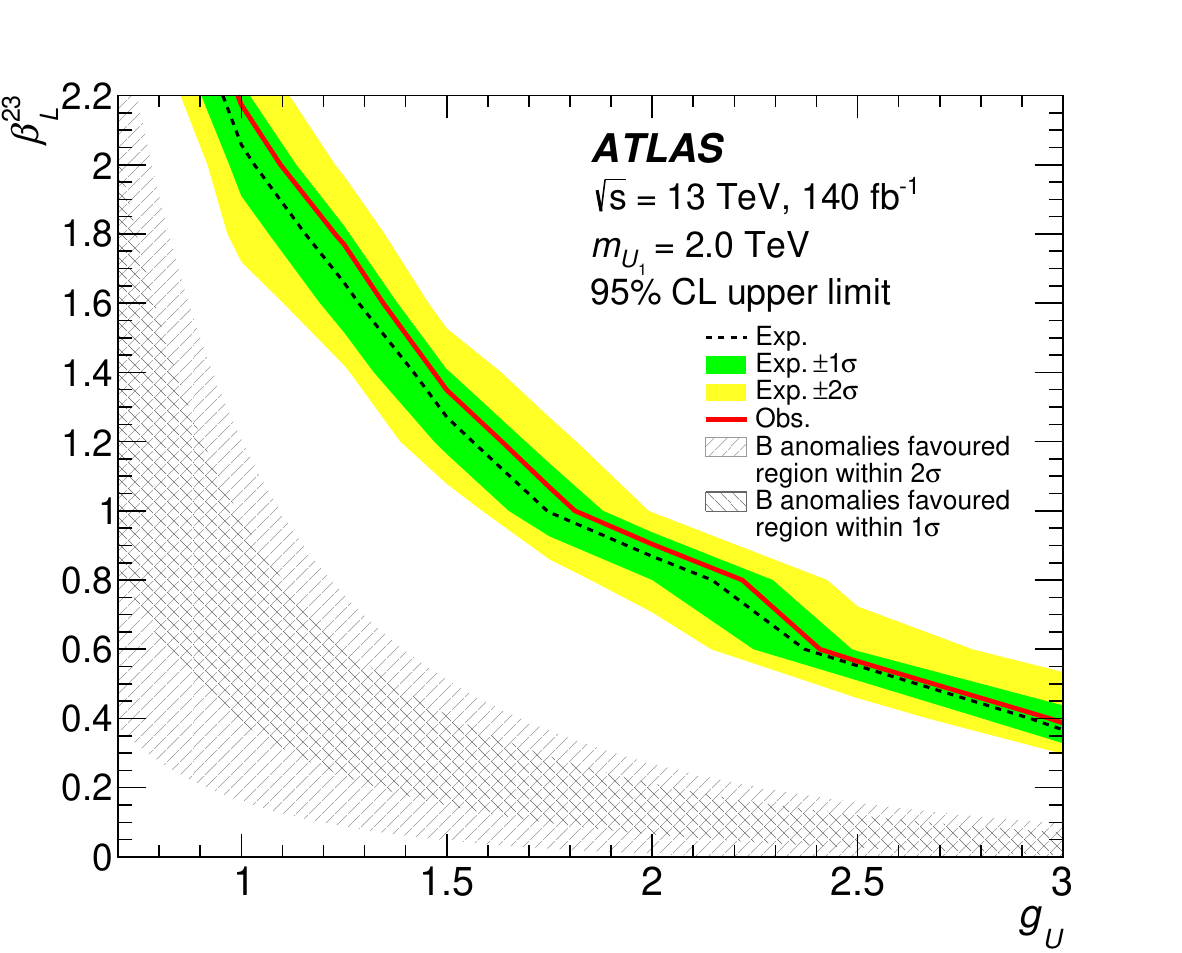}} \\
\subfloat[]{\includegraphics[width=0.45\linewidth]{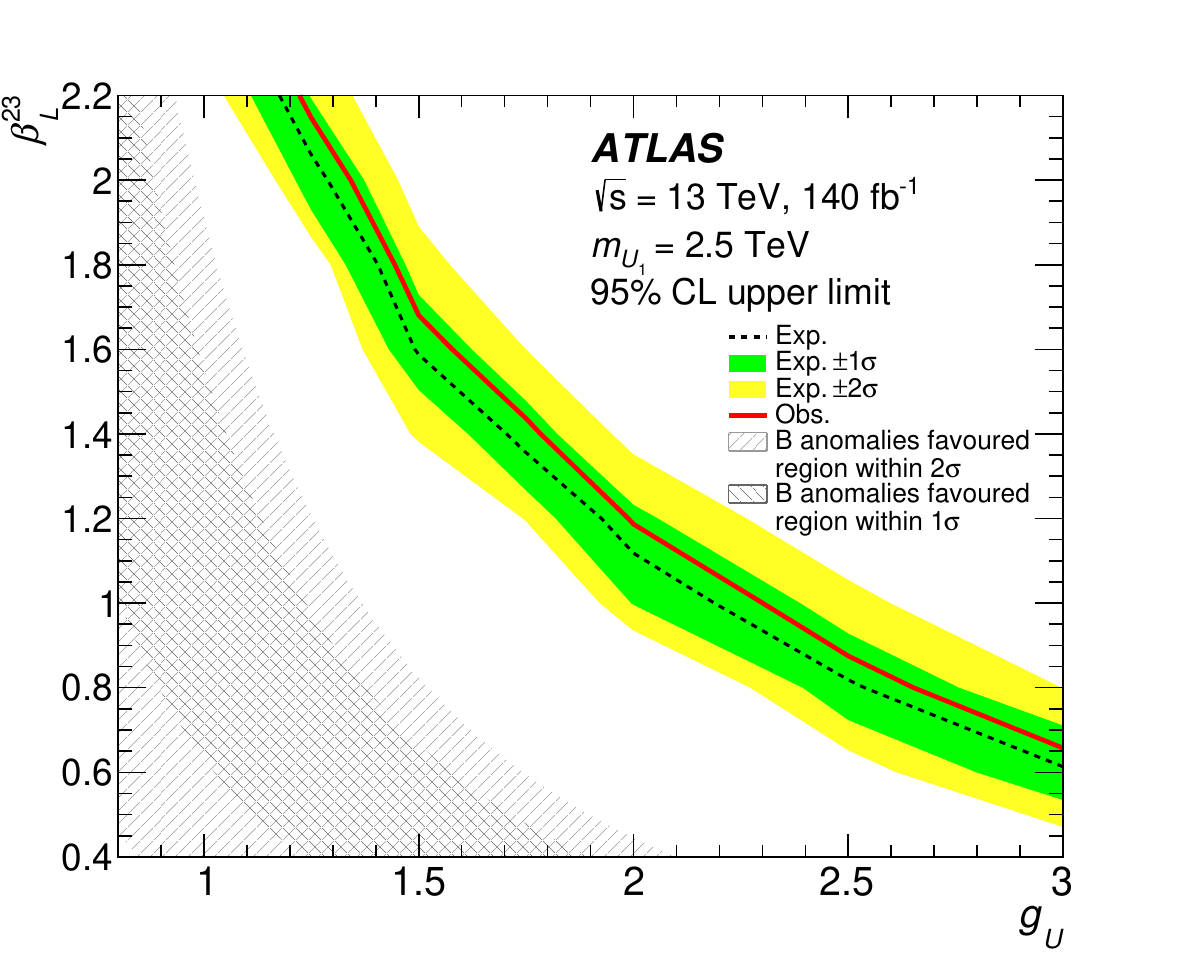}}
\subfloat[]{\includegraphics[width=0.45\linewidth]{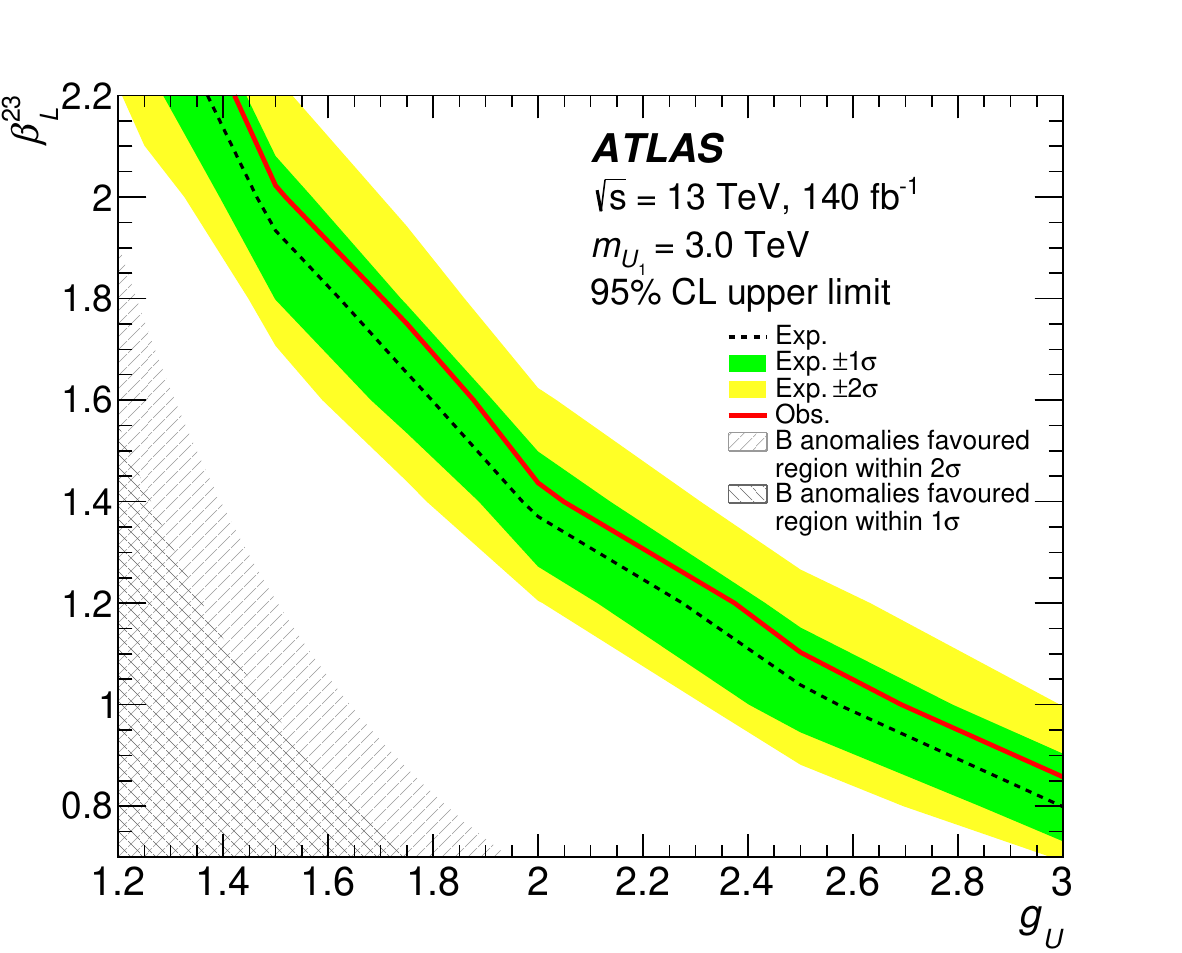}} \\
\caption{Observed (red full line) and expected (black dashed line) 95\% CL upper limits on $\beta_\text{L}^{23}$ as a function of $g_U$ at (a) $\MLQ=1.5~\TeV$, (b) 2~\TeV, (c) 2.5~\TeV and (d) 3.0~\TeV. The shaded regions are favoured by $B$ anomalies at the 1$\sigma$ and 2$\sigma$ levels, converted from $C_\text{LL}^{c}$ limit in Ref.~\cite{Aebischer:2022oqe} using Eq.~(\ref{eq:CLL}).}
\label{fig:unblined_limit_beta_gu}
\end{figure}


%

%
%
\FloatBarrier

\section{Conclusion}
\label{sec:conclusion}
This Letter presents the first ATLAS search targeting high-mass $U_1$ leptoquark signals in the $\tau\nu(+b)$ final state with a hadronically decaying
$\tau$-lepton and missing transverse momentum.  The search uses 140~fb$^{-1}$ of data from 13~TeV proton--proton collisions at the LHC.
The analysis is designed to probe both resonant production and non-resonant $t$-channel exchange of the leptoquark, covering a wide range of coupling scenarios.
No significant excess is observed over the SM background.
Upper limits on the $U_1$ vector-leptoquark couplings are set at 95\% CL for masses between 1.5~TeV and 3.0~TeV.
The results probe part of the parameter space at high $\beta_\text{L}^{23}$ values, which is not covered by the previous $\tau\tau$ searches.

\section*{Acknowledgements}
%

%

%
%

%
%

We thank CERN for the very successful operation of the LHC and its injectors, as well as the support staff at
CERN and at our institutions worldwide without whom ATLAS could not be operated efficiently.

The crucial computing support from all WLCG partners is acknowledged gratefully, in particular from CERN, the ATLAS Tier-1 facilities at TRIUMF/SFU (Canada), NDGF (Denmark, Norway, Sweden), CC-IN2P3 (France), KIT/GridKA (Germany), INFN-CNAF (Italy), NL-T1 (Netherlands), PIC (Spain), RAL (UK) and BNL (USA), the Tier-2 facilities worldwide and large non-WLCG resource providers. Major contributors of computing resources are listed in Ref.~\cite{ATL-SOFT-PUB-2026-001}.

We gratefully acknowledge the support of ANPCyT, Argentina; YerPhI, Armenia; ARC, Australia; BMWFW and FWF, Austria; ANAS, Azerbaijan; CNPq and FAPESP, Brazil; NSERC, NRC and CFI, Canada; CERN; ANID, Chile; CAS, MOST, NSFC and FRFCU, China; Minciencias, Colombia; MEYS CR, Czech Republic; DNRF and DNSRC, Denmark; IN2P3-CNRS and CEA-DRF/IRFU, France; SRNSFG, Georgia; BMFTR, HGF and MPG, Germany; GSRI, Greece; RGC and Hong Kong SAR, China; ICHEP and Academy of Sciences and Humanities, Israel; INFN, Italy; MEXT and JSPS, Japan; CNRST, Morocco; NWO, Netherlands; RCN, Norway; MNiSW, Poland; FCT, Portugal; MNE/IFA, Romania; MSTDI, Serbia; MSSR, Slovakia; ARIS and MVZI, Slovenia; DSI/NRF, South Africa; MICIU/AEI, Spain; SRC and Wallenberg Foundation, Sweden; SERI, SNSF and Cantons of Bern and Geneva, Switzerland; NSTC, Taipei; TENMAK, T\"urkiye; STFC/UKRI, United Kingdom; DOE and NSF, United States of America.

Individual groups and members have received support from BCKDF, CANARIE, CRC and DRAC, Canada; CERN-CZ, FORTE and PRIMUS, Czech Republic; COST, ERC, ERDF, Horizon 2020 and Marie Sk{\l}odowska-Curie Actions, European Union; Investissements d'Avenir Labex, Investissements d'Avenir Idex and ANR, France; DFG and AvH Foundation, Germany; Herakleitos, Thales and Aristeia programmes co-financed by EU-ESF and the Greek NSRF, Greece; BSF-NSF and MINERVA, Israel; NCN and NAWA, Poland; La Caixa Banking Foundation, CERCA and AGAUR programs from Generalitat de Catalunya and PROMETEO and GenT Programmes Generalitat Valenciana, Spain; G\"{o}ran Gustafssons Stiftelse, Sweden; The Royal Society and Leverhulme Trust, United Kingdom; Eric and Wendy Schmidt Fund for Strategic Innovation, United States of America.

In addition, individual members wish to acknowledge support from Chile: Agencia Nacional de Investigaci\'on y Desarrollo (ANID FONDECYT reg. 1230987, FONDECYT 1230812, FONDECYT 1240864, Fondecyt 3240661, Fondecyt Regular 1240721); China: Fundamental Research Funds for the Central Universities (010-63263105), Chinese Ministry of Science and Technology (MOST-2023YFA1605700, MOST-2023YFA1609300), National Natural Science Foundation of China (NSFC 12275265, NSFC-W2543005); Czech Republic: Czech Science Foundation (GACR - 24-11373S), Ministry of Education Youth and Sports (ERC-CZ-LL2327, FORTE CZ.02.01.01/00/22\_008/0004632), PRIMUS Research Programme (PRIMUS/21/SCI/017); EU: H2020 European Research Council (ERC - 101002463); European Union: European Research Council (BARD No. 101116429, ERC - 101219398, ERC - 948254, ERC 101089007), European Regional Development Fund (HE COFUND GA No.101081355, ERDF), Marie Sklodowska-Curie Actions (GAP-101168829); France: Agence Nationale de la Recherche (ANR-21-CE31-0013, ANR-22-EDIR-0002, ANR-24-CE31-0504-01); Germany: Deutsche Forschungsgemeinschaft (DFG - 469666862); China: Research Grants Council (GRF); Italy: Istituto Nazionale di Fisica Nucleare (LHC-MIUR - 28003/2025), Ministero dell'Università e della Ricerca (NextGenEU 153D23001490006 M4C2.1.1, NextGenEU I53D23000820006 M4C2.1.1, NextGenEU I53D23001490006 M4C2.1.1, SOE2024\_0000023); Japan: Japan Society for the Promotion of Science (JSPS KAKENHI  JP25H0063, JSPS KAKENHI JP22H01227, JSPS KAKENHI JP22H04944, JSPS KAKENHI JP22KK0227, JSPS KAKENHI JP24K23939, JSPS KAKENHI JP24KK0251, JSPS KAKENHI JP25H00650, JSPS KAKENHI JP25H01291, JSPS KAKENHI JP25K01011, JSPS KAKENHI JP25K01023); Poland: Polish National Science Centre (NCN OPUS 2023/51/B/ST2/02507, NCN OPUS nr 2022/47/B/ST2/03059, UMO-2019/34/E/ST2/00393, UMO-2021/42/E/ST2/00350, UMO-2022/47/O/ST2/00148, UMO-2023/49/B/ST2/04085, UMO-2023/51/B/ST2/00920, UMO-2024/53/N/ST2/00869); Spain: Agència de Gestió d'Ajuts Universitaris i de Recerca. (AGAUR - 2023 BP 00141), Ministry of Science and Innovation (RYC2019-028510-I, RYC2020-030254-I, RYC2021-031273-I, RYC2022-038164-I), Ministerio de Ciencia, Innovación y Universidades/Agencia Estatal de Investigaci\'on (EU NextGenerationEU (PRTR-C17.I1), PID2022-142604OB-C22); Sweden: Carl Trygger Foundation (Carl Trygger Foundation CTS 22:2312), Swedish Research Council (Swedish Research Council 2023-04654, VR 2021-03651, VR 2022-03845, VR 2022-04683, VR 2023-03403, VR 2024-05451, VR 2025-05940), Knut and Alice Wallenberg Foundation (KAW 2023.0366); Switzerland: Swiss National Science Foundation (SNSF - PCEFP2\_194658); United Kingdom: The Binks Trust, Royal Society (NIF-R1-231091); United States of America: U.S. Department of Energy (ECA DE-AC02-76SF00515), John Templeton Foundation (John Templeton Foundation 63206), Neubauer Family Foundation.

%
%


%
%
%
%
%
%

%
%

%
%
\printbibliography
\clearpage
\input{atlas_authlist}
%

%
%

%

%
%
%
%

\end{document}

%% file: atlas_authlist.tex
 
\begin{flushleft}
\hypersetup{urlcolor=black}
{\Large The ATLAS Collaboration}

\bigskip

\AtlasOrcid[0000-0002-6665-4934]{G.~Aad}$^\textrm{\scriptsize 102}$,
\AtlasOrcid[0000-0001-7616-1554]{E.~Aakvaag}$^\textrm{\scriptsize 17}$,
\AtlasOrcid[0000-0002-5888-2734]{B.~Abbott}$^\textrm{\scriptsize 121}$,
\AtlasOrcid[0000-0002-0287-5869]{S.~Abdelhameed}$^\textrm{\scriptsize 83b}$,
\AtlasOrcid[0000-0002-1002-1652]{K.~Abeling}$^\textrm{\scriptsize 54}$,
\AtlasOrcid[0000-0001-5763-2760]{N.J.~Abicht}$^\textrm{\scriptsize 48}$,
\AtlasOrcid[0000-0002-8496-9294]{S.H.~Abidi}$^\textrm{\scriptsize 30}$,
\AtlasOrcid[0009-0003-6578-220X]{M.~Aboelela}$^\textrm{\scriptsize 44}$,
\AtlasOrcid[0000-0002-9987-2292]{A.~Aboulhorma}$^\textrm{\scriptsize 36e}$,
\AtlasOrcid[0000-0001-5329-6640]{H.~Abramowicz}$^\textrm{\scriptsize 154}$
% ...
\bigskip
\\

% ...
$^{17}$Department for Physics and Technology, University of Bergen, Bergen; Norway.\\
% ...
$^{30}$Physics Department, Brookhaven National Laboratory, Upton NY; United States of America.\\
% ...
$^{36}$$^{(a)}$Facult\'e des Sciences Ain Chock, Universit\'e Hassan II de Casablanca;$^{(b)}$Facult\'{e} des Sciences, Universit\'{e} Ibn-Tofail, K\'{e}nitra;$^{(c)}$Facult\'e des Sciences Semlalia, Universit\'e Cadi Ayyad, LPHEA-Marrakech;$^{(d)}$LPMR, Facult\'e des Sciences, Universit\'e Mohamed Premier, Oujda;$^{(e)}$Facult\'e des sciences, Universit\'e Mohammed V, Rabat;$^{(f)}$Institute of Applied Physics, Mohammed VI Polytechnic University, Ben Guerir; Morocco.\\
% ...
$^{44}$Physics Department, Southern Methodist University, Dallas TX; United States of America.\\
% ...
$^{48}$Fakult\"{a}t Physik, Technische Universit{\"a}t Dortmund, Dortmund; Germany.\\
% ...
$^{54}$II. Physikalisches Institut, Georg-August-Universit\"{a}t G\"ottingen, G\"ottingen; Germany.\\
% ...
$^{83}$$^{(a)}$Khalifa University of Science and Technology, Abu Dhabi;$^{(b)}$New York University Abu Dhabi, Abu Dhabi;$^{(c)}$United Arab Emirates University, Al Ain;$^{(d)}$University of Sharjah, Sharjah; United Arab Emirates.\\
% ...
$^{102}$CPPM, Aix-Marseille Universit\'e, CNRS/IN2P3, Marseille; France.\\
% ...
$^{121}$Homer L. Dodge Department of Physics and Astronomy, University of Oklahoma, Norman OK; United States of America.\\
% ...
$^{154}$Raymond and Beverly Sackler School of Physics and Astronomy, Tel Aviv University, Tel Aviv; Israel.\\
% ...
$^{a}$ Also at Affiliated with an institute formerly covered by a cooperation agreement with CERN.\\
$^{b}$ Also at An-Najah National University, Nablus; Palestine.\\
$^{c}$ Also at Borough of Manhattan Community College, City University of New York, New York NY; United States of America.\\
$^{d}$ Also at Center for Interdisciplinary Research and Innovation (CIRI-AUTH), Thessaloniki; Greece.\\
$^{e}$ Also at Centre of Physics of the Universities of Minho and Porto (CF-UM-UP); Portugal.\\
$^{f}$ Also at CERN, Geneva; Switzerland.\\
% ...

\end{flushleft}
